\documentclass[%
 aip,
 amsmath,amssymb,
 reprint,%
]{revtex4-1}

\usepackage{graphicx}
\usepackage{amsmath}
\usepackage{dcolumn}
\usepackage{bm}
\usepackage[usenames]{color}

\begin{document}

\preprint{AIP/123-QED}

\title[]{Extending the applicability of an open-ring trap to perform experiments with a single laser-cooled ion}

\author{J.M.~Cornejo}
\affiliation{Departamento de F\'isica At\'omica, Molecular y Nuclear, Universidad de Granada, 18071, Granada, Spain}
\author{M.~Colombano}
\affiliation{Departamento de F\'isica At\'omica, Molecular y Nuclear, Universidad de Granada, 18071, Granada, Spain}
\author{J.~Dom\'enech}
\affiliation{Departamento de F\'isica At\'omica, Molecular y Nuclear, Universidad de Granada, 18071, Granada, Spain}
\author{M.~Block}
\affiliation{GSI Helmholtzzentrum f\"ur Schwerionenforschung GmbH, 64291, Darmstadt, Germany}
\affiliation{Helmholtz-Institut Mainz, 55099,  Mainz, Germany}
\affiliation{Institut f\"ur Kernchemie, University of Mainz, 55099 Mainz, Germany}
\author{P.~Delahaye}
\affiliation{Grand Acc\'el\'erateur National d'Ions Lourds, 14000, Caen, France}
\author{D.~Rodr\'iguez}
\email[]{Author to whom correspondence should be addressed: danielrodriguez@ugr.es}

\affiliation{Departamento de F\'isica At\'omica, Molecular y Nuclear, Universidad de Granada, 18071, Granada, Spain}

\date{\today}

\begin{abstract}

An open-ring ion trap, also referred to as transparent trap was initially built up to perform $\beta$-$\nu$ correlation experiments with radioactive ions. This trap geometry is also well suited to perform experiments with laser-cooled ions, serving for the development of a new type of Penning trap, in the framework of the project TRAPSENSOR at the University of Granada. The goal of this project is to use a single $^{40}$Ca$^+$ ion as detector for single-ion mass spectrometry. Within this project and without any modification to the  initial electrode configuration, it was possible to perform Doppler cooling on $^{40}$Ca$^+$ ions, starting from large clouds and reaching single ion sensitivity. This new feature of the trap might be important also for other experiments with ions produced at Radioactive Ion Beam (RIB) facilities. In this publication, the trap and the laser system will be described, together with their performance with respect to laser cooling applied to large ion clouds down to a single ion. 

\end{abstract}

\maketitle

\section{Introduction}

Laser cooling of stable ions stored in electromagnetic traps is a field of research which started a few decades ago and it is nowadays very active \cite{Leib2003}. This technique is prominently applied in precision laser-spectroscopy  \cite{Chwa2009}, for improving the accuracy and precision in optical clocks \cite{Schm2005}, and for quantum simulations and quantum information processing \cite{blat2012}. In all these cases, only one non-decaying (stable) ion or a chain comprising a few ones is stored in a radiofrequency (RF) linear Paul trap, or even transported between several of these devices \cite{Kiel2002}. Experiments with Penning traps are less common \cite{Mitc1998,Mada2013}. They are also carried out with stable ions or to use a coolant ion to perform sympathetic cooling on other ion species not accessible with lasers \cite{Ande2013}. Despite these many applications, laser cooling has not been applied up to now in precision Penning-trap mass spectrometry (PTMS) \cite{Blau2013}.\\

\noindent The work presented in this paper refers to developments carried out at the University of Granada in the framework of the project TRAPSENSOR. The goal is to perform single trapped-ion mass-spectrometry based on optical detection using a laser-cooled $^{40}$Ca$^+$ ion, referred to as {\it Quantum Sensor} (QS) \cite{Rodr2012}, in order to increase the sensitivity of the Penning-trap method. Up to date, the highest sensitivity, i.e. measuring only with single ion, has been reached on stable ions with low or medium mass-to-charge ratios using electronic detection (e.g. \cite{Reds2008}). The QS technique would be favorable for a PTMS system such as SHIPTRAP at GSI Darmstadt, where high-precision mass measurements of heavy nuclides ($Z>100$) are performed \cite{Bloc2005,Bloc2010}. The technique under development relies on the communication of two ions, proposed in 1990 by D.J. Heinzen and D.J. Wineland \cite{Wine1990}, which has not been demonstrated up to date with ions stored in different traps. Communication and manipulation at the quantum level has been proven for ions from different species stored in the same trap \cite{Schm2005}, and with ions of the same species stored in the same trap, but confined in different DC potential wells \cite{Brow2011}. All these experiments have been performed in linear traps, while for the project TRAPSENSOR, a trap with cylindrical symmetry has to be used. The so-called open-ring trap or transparent Paul trap, meets the requirements with respect to symmetry, allowing also for access of the laser beams and for the collection of the fluorescence photons in the way described in the initial proposal \cite{Rodr2012}. The configuration adopted as Paul trap is identical to the one to be used as a Penning trap. Only the energy levels for the $^{40}$Ca$^+$ ion will change due to the Zeeman splitting in the high magnetic field of the device, which will involve more laser beams.\\

\begin{figure*} 
\hspace{0cm}
\includegraphics[width=0.8\textwidth]{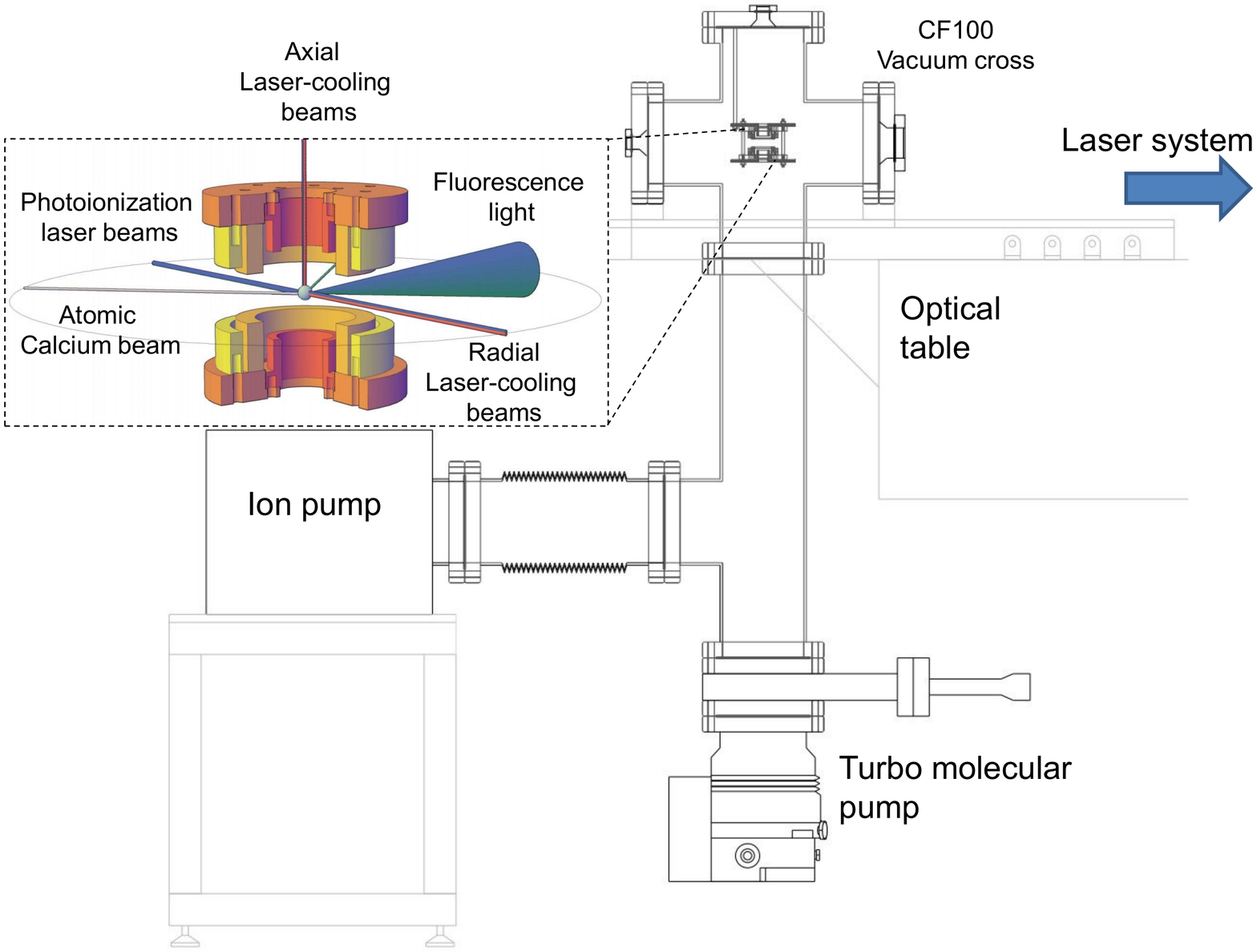}
\caption{(Color online) 2D-CAD drawing of the vacuum chamber containing the transparent Paul trap. The inset shows a 3D cut of the trap indicating the laser beams used to photoionize the calcium atoms and those to perform laser cooling. The vacuum chamber and tube are fixed at the end of an optical table where the lasers and optics are installed. The 6-way CF100 vacuum cross has another four CF16 ports located in the radial plane, which serve to place the atomic-beam source, an electron gun and a Faraday cup detector. These ports are not shown in the figure. The electron beam has been also utilized to ionize the calcium atoms in the center of the trap.\label{fig:1.2}}
\end{figure*}

\noindent The open-ring trap utilized for this work is a large device initially developed for $\beta$-$\nu $ correlation experiments using ions produced at the SPIRAL (ISOL-type)  source at the Grand Acc\'el\'erateur National d`Ions Lourds (GANIL) \cite{Rodr2006,Ban2013,Dela2015}. The device has already yielded important scientific results in nuclear and atomic physics \cite{Flec2011,Cour2012,Cour2013} and might still be used for experiments like for instance tests of the Standard Model involving the $\beta$-decay of polarized $^{23}$Mg ($T_{1/2}=11.3$~s) and $^{39}$Ca ($T_{1/2}=860$~ms) ions for $D$-correlation measurements, following the determination of the so-called $a_{\beta \nu}$ parameter in nuclear $\beta $ decay \cite{Dela2015}, or for the determination of the $V_{ud}$ element of the Cabibbo-Kobayashi-Maskawa (CKM) matrix from mirror decays \cite{Navi2009}. Furthermore, laser-cooled ions might be used to cool sympathetically the radioactive ions down to sub-Kelvin temperatures. It has been shown that the main contribution to the systematic uncertainty in the determination of $a_{\beta \nu}$ from the decay of $^6$He$^+$ ions, is the size of the ion cloud \cite{Flec2011}.\\

\noindent In the following, the setup around the open-ring trap built in the framework of the project TRAPSENSOR, to perform laser cooling on  $^{40}$Ca$^+$ is described in detail. The characterization of this device with respect to laser cooling is also discussed. Relevant aspects like storage time, cooling time, capacity, and sensitivity of the trap after applying Doppler cooling will be discussed. The importance of these features for experiments with stable and radioactive ions will be addressed, demonstrating the versatility of the device.

\section{The experimental set-up}

\begin{figure}
\hspace{0cm}
\includegraphics[width=0.5\textwidth]{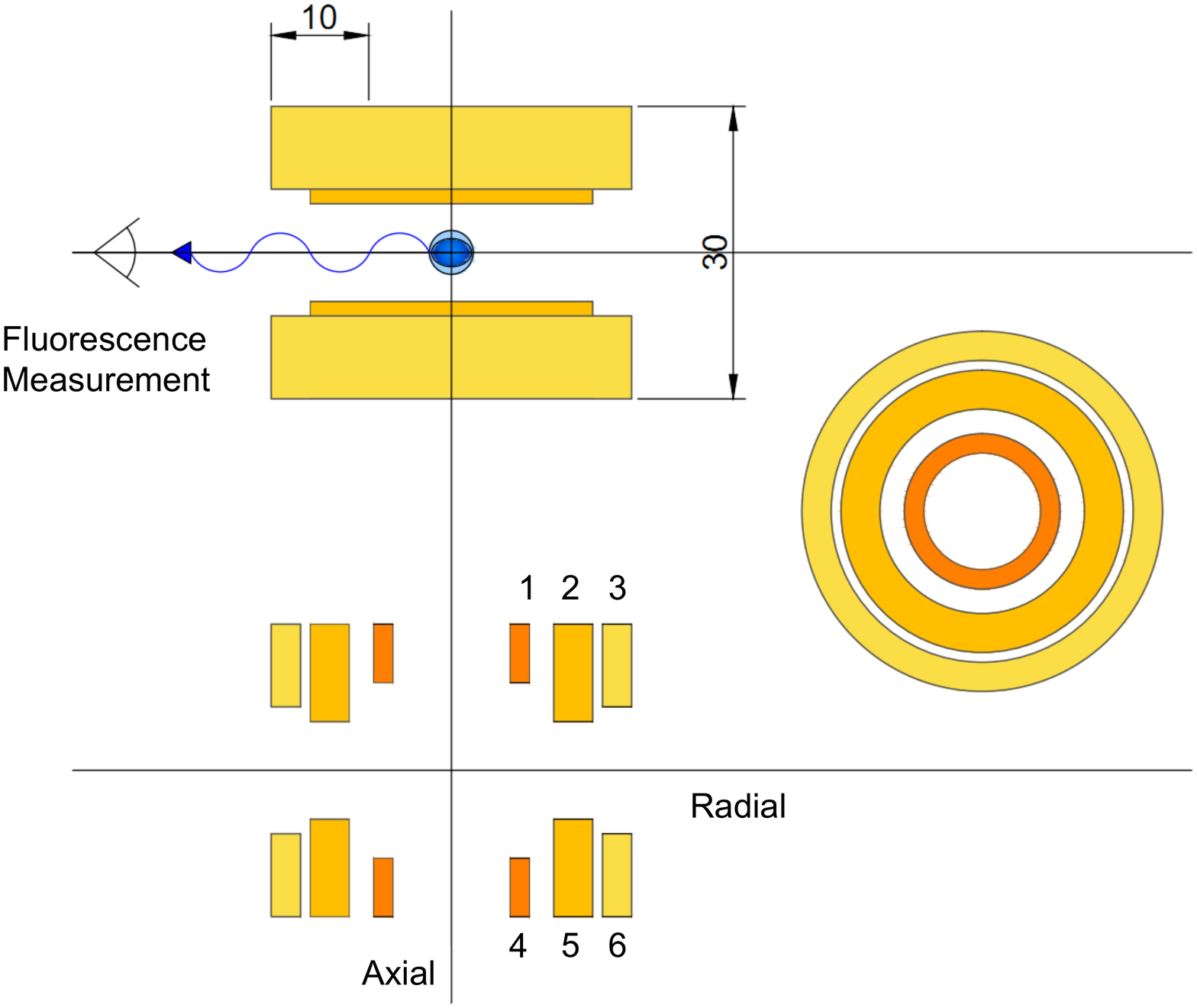}
\vspace{0cm}
\caption{(Color online) Different views of the ion trap shown in Fig.~\ref{fig:1.2}. The RF voltage is applied to rings 1 and 4. Rings 2 and 5 are connected to a DC-voltage power supply, and rings 3 and 6 are grounded for the measurements presented here. \label{fig:1.3}}
\end{figure}

The electrode configuration for the open-ring trap built up at the University of Granada is identical to the one previously installed at GANIL \cite{Ban2013}. It is now used in a different experimental arrangement, to perform Doppler cooling on $^{40}$Ca$^+$ ions. A 2D-CAD drawing of the trap inside a 6-way CF100 vacuum cross, attached to an optical table is shown in Fig.~\ref{fig:1.2}. The chamber was initially pumped by a turbomolecular pump backed by a roughing pump. Thereafter, a 300~l/s ion pump was switched on, enabling to reach a final pressure of $1.5\times 10^{-10}$~Torr at the pump side, for these experiments. The open-ring trap, consisting of a pair of three rings, is zoomed inside the dashed box (Fig.~\ref{fig:1.2}). Figure~\ref{fig:1.3} shows three different views of the trap. The trapping quadrupolar field in the center of the trap is generated by applying a radiofrequency (RF) voltage to the inner electrodes (numbered 1 and 4 in Fig.~\ref{fig:1.3}). Electrodes numbered 2 and 5 are connected to a stable DC-voltage power supply and the outer ones are grounded.\\ 

\begin{figure*}
\hspace{0cm}
\includegraphics[width=1\textwidth]{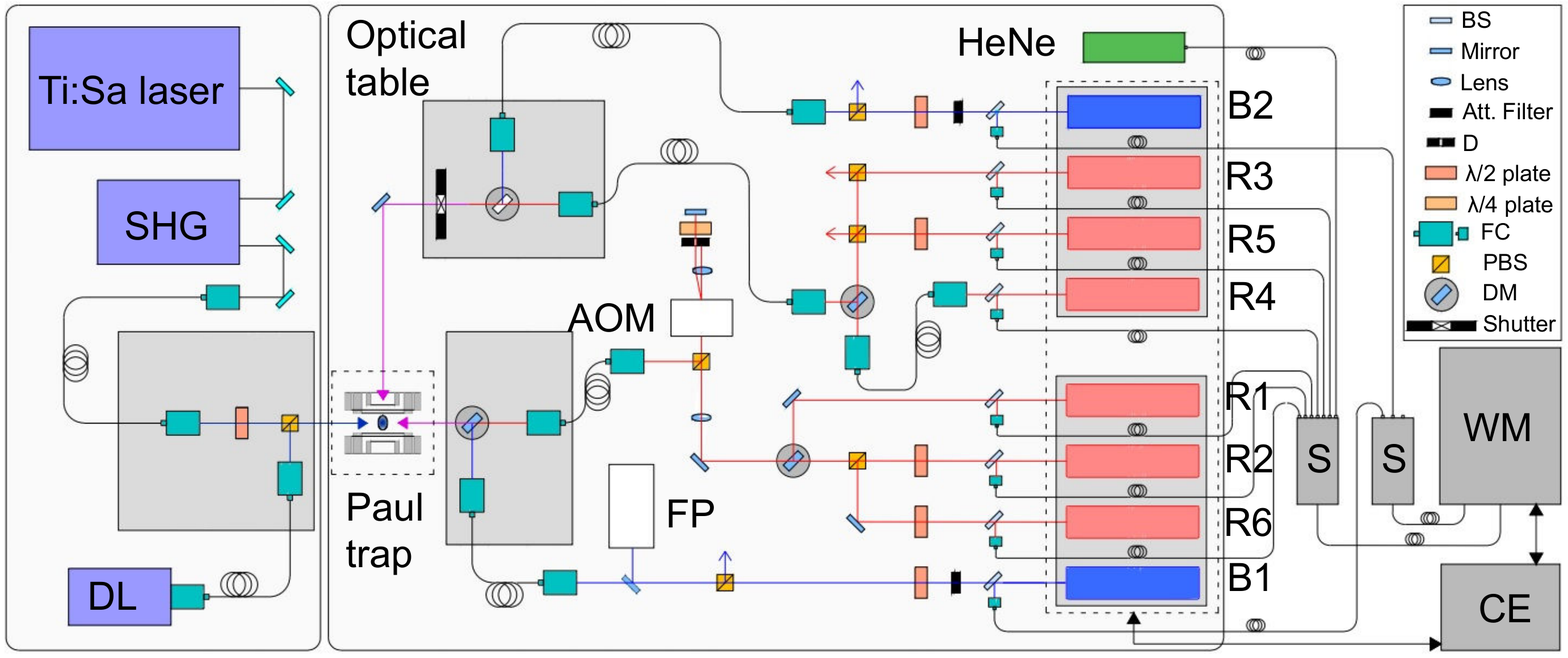}
\vspace{-5.5cm}
\caption{(Color online) Sketch of the optical table with the laser system built for the experiment. This system serves also for laser cooling in a Penning trap. SHG stands for Second Harmonic Generation, DL for fixed-frequency diode laser, AOM for Acoustic Optical Modulator, FP for Fabry-Perot Interferometer, B and R for blue and red (tunable diode lasers), CE for control electronics, WM for wavelength meter, and S for switch.  Minor components are labeled in the box on the upper right side of the figure. BS stands for Beam Splitter, D for Diaphragm, FC for Fiber Coupler, PBS for Polarizing Beam Splitter, and DM for Dichroic Mirror. Further details are given in the text. \label{fig:1.1}}
\end{figure*}

\noindent The ions are created inside the trap by two step photoionization of calcium atoms delivered after heating a commercial oven. A Titanium Sapphire (Ti:Sa) laser (Sirah Matisse TX) and a  Second Harmonic Generation (SHG) system from Toptica, serves to produce a laser beam with the frequency of the 4s$^2$S$_{1/2}\rightarrow$4s$^2$P$_{3/2}$ transition in the $^{40}$Ca atom. The ionization is accomplished by superimposing 375~nm light from a compact diode laser (Toptica ibeam smart) for the second photoionzation step. The Ti:Sa laser and the SHG system are located in an optical table with dimensions 2.4~m x 1.5~m, close to the trap setup. They are sketched with reduced dimensions besides the main optical table in Fig.~\ref{fig:1.1}.\\

\noindent The laser system used for Doppler cooling comprises eight external cavity diode lasers in Littrow configuration (ECDL) from Toptica (DL pro) with tunable wavelengths around 397~nm (x2), 866~nm (x4) and 854~nm (x2). Besides the cooling transition at 397~nm (4s$^2$S$_{1/2}\rightarrow$4s$^2$P$_{3/2}$), and the repumping laser at 866~nm (3d$^2$D$_{3/2}\rightarrow$4s$^2$P$_{3/2}$), the 854~nm laser (3d$^2$D$_{5/2}\rightarrow$4s$^2$P$_{3/2}$), was necessarily added in the radial direction, also for repumping. Figure~\ref{fig:1.1} shows schematically the system on an optical table (2.4~m x 1.5~m). Two laser-beam lines have been used in the experiments, one in the radial and the other in the axial direction of the trap (Fig.~\ref{fig:1.2}). The trap axes are oriented in a different way compared to the one shown in Fig.~\ref{fig:1.1}. In the experimental setup, the axial direction of the trap is perpendicular to the optical table. The profiles of the lasers in the radial direction, as well as the overlap between them, were measured using a beam-monitor system (not shown in the figure), located at a distance from the fiber-coupler outputs, equal to that from the trap center.  The 397-nm laser has an elliptical shape and the orthogonal projections have {\it FWHM} of 440 and 525~$\mu$m, respectively. The 866 and 854-nm lasers have circular profiles and the {\it FWHM} of the projections are 1.5~mm. The laser power will be discussed together with the experimental results.\\

\noindent The lasers are locked to a wavelength meter (High Finesse WSU), which has an absolute accuracy of 10~MHz (3$\sigma$), and is calibrated using the 632~nm transition from a stabilized HeNe laser. A small fraction of the output from each ECDL is coupled by an optical fiber to a switch, which delivers the light from the different lasers sequentially into the wavemeter. There is one switch for the UV lasers (B1 and B2 in Fig.~\ref{fig:1.1}) and another for the  Infrared (IR) ones (R1 to R6 in Fig.~\ref{fig:1.1}). PID regulation is performed using the wavelength measurement and generating a voltage signal through a 14-bit analogue card with a resolution of 0.5~mV, which is fed into the piezo-voltage regulation system of the laser in the control electronics box. The performance of this system was already shown \cite{Corn2014} for the eight lasers, as they will be requested to perform Doppler cooling in a Penning trap. The stability of the lasers used for the experiments presented in this paper is shown for one of the measurements in Tab.~\ref{Stability}. The 866~nm lasers can be switched ON and OFF by means of an acoustic optical modulator (radial laser beam), or a shutter valve (axial laser beam).\\

\begin{table}
\caption{Standard deviations (bandwidths) in MHz of the laser frequencies utilized in the experiments described in this paper (Fig.~\ref{fig:1.1}). These deviations are taken from one of the runs, where all frequencies were locked without being scanned. The directions of the laser beams with respect to the trap are also indicated. \label{Stability}}
\begin{ruledtabular}. 
\begin{tabular}{cccccc}
$\sigma $(B1) &$\sigma $(B2) &$\sigma $(R1) & $\sigma $(R2) & $\sigma $(R5) &$\sigma $(R6) \\
axial & radial& radial& radial& axial &radial\\
\hline
\\
5.5 & 1.2 & 3.4& 1.3& 1.0& 1.8\\

\end{tabular}
\end{ruledtabular}
\end{table}

\noindent The fluorescence light from the cooling transition (397~nm) is collected with an EMCCD (Electron Multiplier Charged Couple Device) from Andor (IXON3). The EMCCD sensor comprises 512 x 512 pixels. This camera is attached to an optical system, designed to collimate the fluorescence light from the trap center to the EMCCD sensor. The system consists of two lenses with focal length $f= 100$~mm, and two doublets (Thorlabs  MAP1030100-A and MAP103075-A). It is located behind an inverted view port, and oriented 90~degrees with respect to the laser-cooling beams, in the radial plane. The system provides in all, a magnification ranging from $\approx 8.25$ to $\approx 9.2$, so that the effective size of the EMCCD sensor for the experiments reported here is slightly below 1~mm$^2$. An interference filter with a bandwidth of about 5~nm is located in front of the EMCCD to collect only photons with wavelength around 397~nm. \\

\noindent The acquisition system has been developed in LABVIEW. It is running in several computers connected in a local network. The system allows scanning the lasers following different functions (see Fig.~\ref{fig:1}), the function generator and power for the RF driving field, the DC voltages to run the electronics (oscillator and attenuator) for the AOM, and to monitor and record the fluorescence image and the photon count-rate from the EMCCD, together with all laser frequencies \cite{Esco2014}. 

\section{Laser-cooling of ion clouds}

\begin{figure}
\hspace{-0.5cm}
\includegraphics[width=0.5\textwidth]{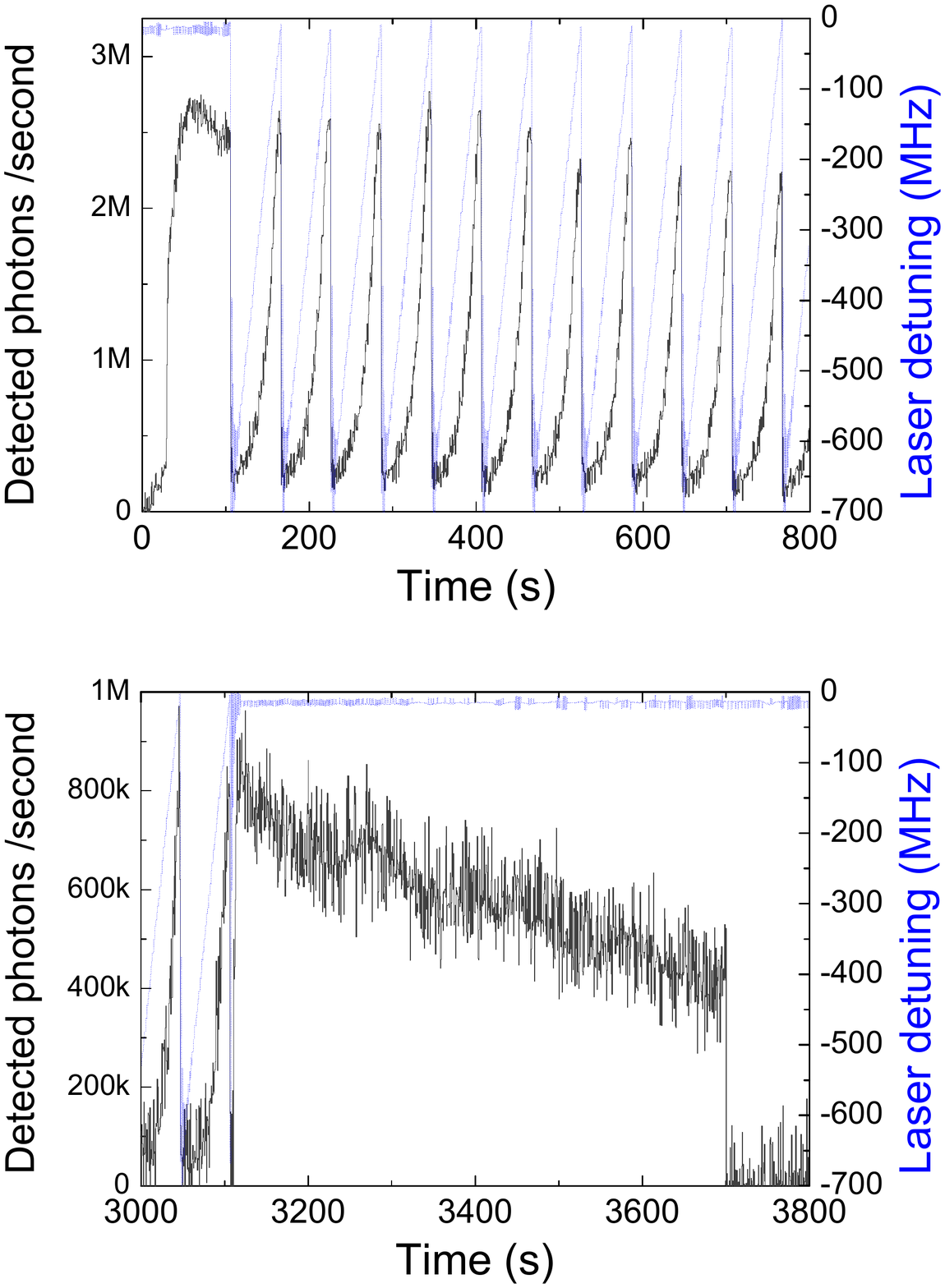}
\vspace{-0.9cm}
\caption{(Color online) Detected photons from a laser-cooled $^{40}$Ca$^+$ ion cloud, collected versus time as a function of the B2-laser detuning from resonance (axis on the right side). Top: 700-seconds time interval after loading. Bottom: Detected signal after 3000~s. For further details see text. \label{fig:1}}
\end{figure}

Figure~\ref{fig:1} shows a long-time measurement of a fluorescence signal from a laser-cooled ion cloud as a function of the detuning of the 397-nm radial-laser beam (B2 in Fig.~\ref{fig:1.1}). The mean value of the background signal was substrated to the total signal. The rest of the lasers including B1 are not scanned, since there is no difference by scanning the two 397-nm lasers (radial and axial) simultaneously. The maximum frequency-detuning in this picture is 700~MHz below resonance, which corresponds to an ion temperature of approximately 370~K and thus, to an ion energy of about 30~meV. The storage time (time constant from the exponential decay) is $\tau = 971(18)$~s. This constant is obtained with the data shown in the lower part of Fig.~\ref{fig:1} after stopping the frequency scan.\\ 

\noindent The open-ring trap can be characterized, as any other Paul trap, by the so-called Mathieu parameter given by \cite{Daw1995}

\begin{equation}\label{eq:qz}
q_u=\frac{eV_\mathrm{RF}}{m\omega_\mathrm{RF}^2u_0^{2}},
\end{equation}

\noindent where $u$ stands for $r$ (radial) and $z$ (axial) direction, $m/e$ is the mass-to-charge ratio of the ion in the trap, $V_{\scriptsize{\hbox{RF}}}$ is the amplitude of the RF voltage applied to perform the quadrupolar potential, and $\omega_{\scriptsize{\hbox{RF}}}$ is $2\pi \nu_{\scriptsize{\hbox{RF}}}$ with $\nu_{\scriptsize{\hbox{RF}}}$ the frequency. For the measurements presented in this publication, $V_{\scriptsize{\hbox{RF}}}=111.5$~V  (223~V$_{\scriptsize{\hbox{pp}}}$) and $\nu_{\scriptsize{\hbox{RF}}}=650$~kHz. This corresponds to $q_r=0.205$ and $q_z=0.410$. These values are obtained from the axial oscillation frequency of the ion

\begin{equation}
\omega _z^{\scriptsize{\hbox{sec}}}=\frac{q_z}{2\sqrt{2}}\omega _{\scriptsize{\hbox{RF}}},
\end{equation}

\noindent after applying an external dipolar field in resonance with it. The $q_z$-value of 0.410 ($\nu _z=94$~kHz) is in the limit of the so-called adiabatic approximation and thus, no strong heating effects from the driving field on the ions are expected in the cooling process.\\

\begin{figure}
\hspace{-0.7cm}
\includegraphics[width=0.52\textwidth]{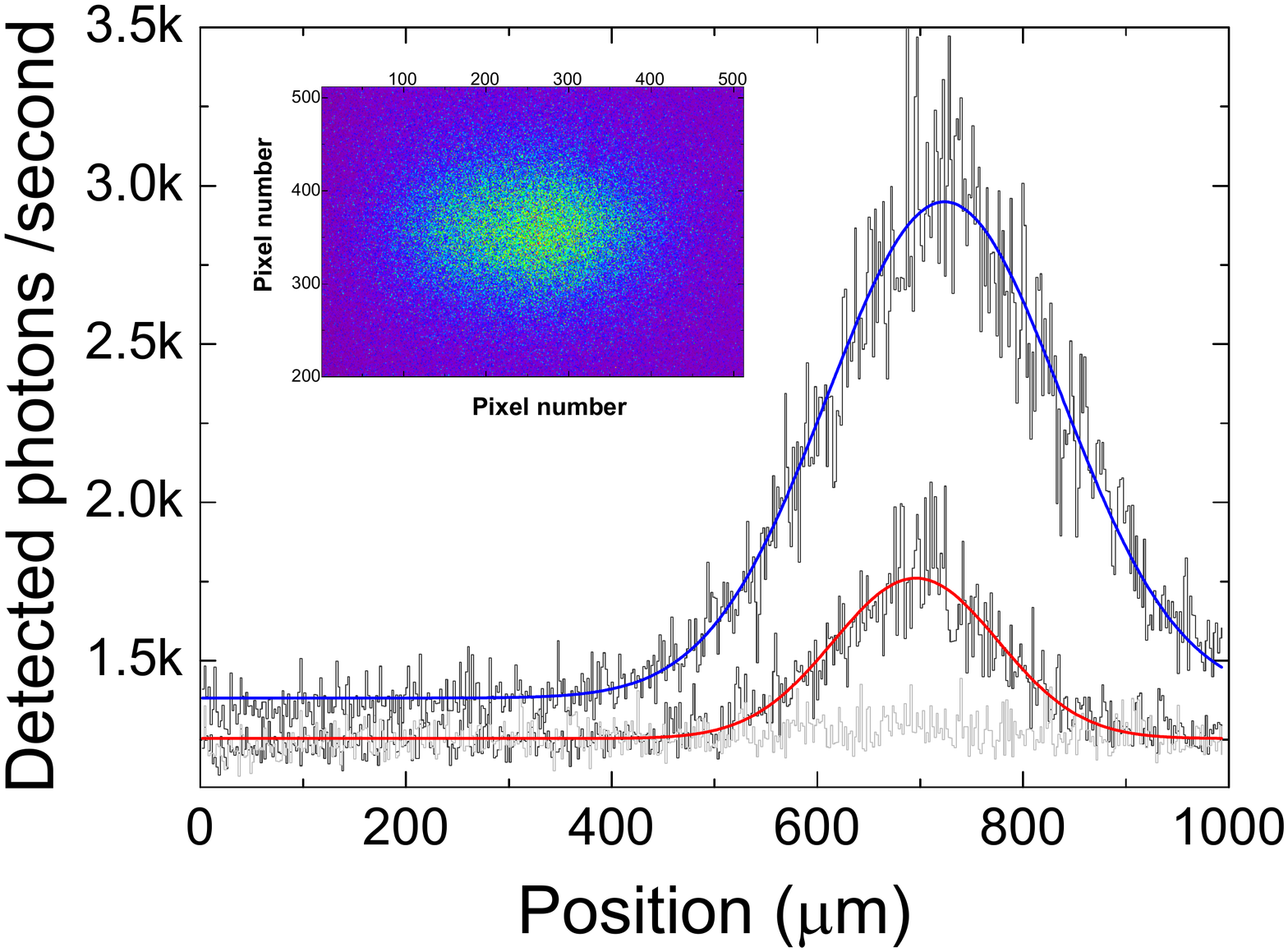}
\vspace{-0.3cm}
\caption{(Color online) Axial distribution for a large (Gaussian fit in blue) and a medium size (Gaussian fit in red) laser-cooled $^{40}$Ca$^+$ ion-cloud. The {\it FWHM} is 269(4) and 189(5)~$\mu$m for the large and the medium size cloud, respectively.  The frequency of the B2 laser was scanned (similarly as it was done in Fig.~\ref{fig:1}), and the image was collected at the maximum fluorescence. The frequencies of the other lasers were not scanned. The light grey histogram represents the background. The inset is an image of the laser-cooled ion cloud. \label{fig:2}}
\end{figure}

\begin{figure}
\hspace{-0.7cm}
\includegraphics[width=0.5\textwidth]{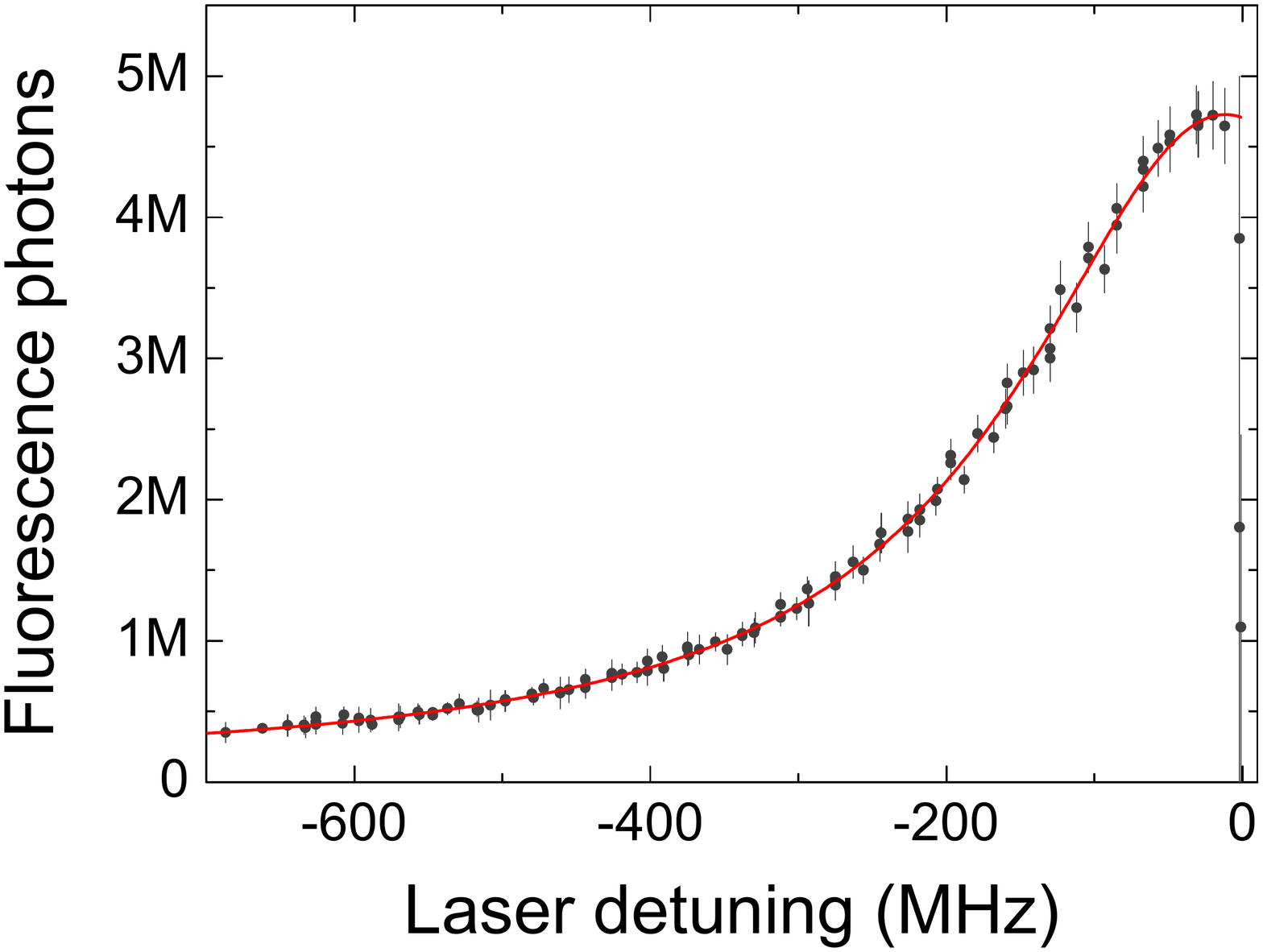}
\vspace{-0.3cm}
\caption{Fluorescence photons collected as a function of the B2-laser detuning from resonance. The  histogram is the average of seven measurements, and corresponds to the largest area one can observe with the optical system. Based on the fluorescence photons collected from a single laser-cooled ion, the number of ions in the cloud might be between 100 and 1000. The data are fitted using a Voigt function (see text for details).\label{fig:new}}
\end{figure}

\noindent Figure~\ref{fig:2} shows the axial projection from two fluorescence images and a Gaussian fit to the data. The light-grey histogram is the background signal. The inset of the figure shows a fluorescence image in the sensor of the EMCCD camera for the larger-size ion cloud. Due to the optical system and the sensor area, the observation of the full fluorescence image from very large ion clouds containing above $\approx 1000$~ions will not be possible. Larger clouds have been observed with a different optical setup in front of the EMCCD camera consisting of two lenses with focal distances of 50 and 78~mm, respectively, and a doublet (Thorlabs MAP103075-A), thus magnifying the image by a factor of $\approx 4$.\\

\noindent It is possible to determine the maximum energy of the ions in the cloud using the pseudopotential model which reads in the axial direction

\begin{equation}\label{eq:3.2.4}
\mathcal{V}(z)=\frac{e V_\mathrm{RF}^{2}}{4mz_0^{4}\omega_\mathrm{RF}^{2}}z^{2}.
\end{equation}

\noindent Applying the Virial theorem, where the mean kinetic energy is half of the total ion-energy in a harmonic potential, one obtains the relationship

\begin{equation}\label{eq:virial}
{E_\mathrm{kin}}=\frac{1}{2}e\mathcal{V}(A_z)=\frac{1}{2}k_BT,
\end{equation}

\noindent in one dimension ($3/2\cdot k_BT$ in three degrees of freedom). The temperature is  given by

\begin{equation}\label{eq:3.2.6}
T=\frac{e^{2}V_\mathrm{RF}^{2}}{4k_B mz_0^{4}\omega_\mathrm{RF}^{2}}A_z^{2},
\end{equation}

\noindent where $A_z$ is half of the semi-axis of the ellipse, and the other parameters are defined in previous formulae. This results in $E_{\scriptsize{\hbox{max}}}\approx 15$~meV for the large cloud presented in the inset of Fig.~\ref{fig:2}. Observation of fluorescence without cooling was only possible for very large ion clouds. In such case, an estimated upper limit for the temperature of about 1200~K was obtained from

\begin{equation}\label{eq:temperature}
T=\frac{mc^{2}}{8k_B \ln 2}\left | \frac{\Delta\nu (D)}{\nu_0}\right |^{2},
\end{equation}

\begin{figure}
\hspace{0cm}
\includegraphics[width=0.5\textwidth]{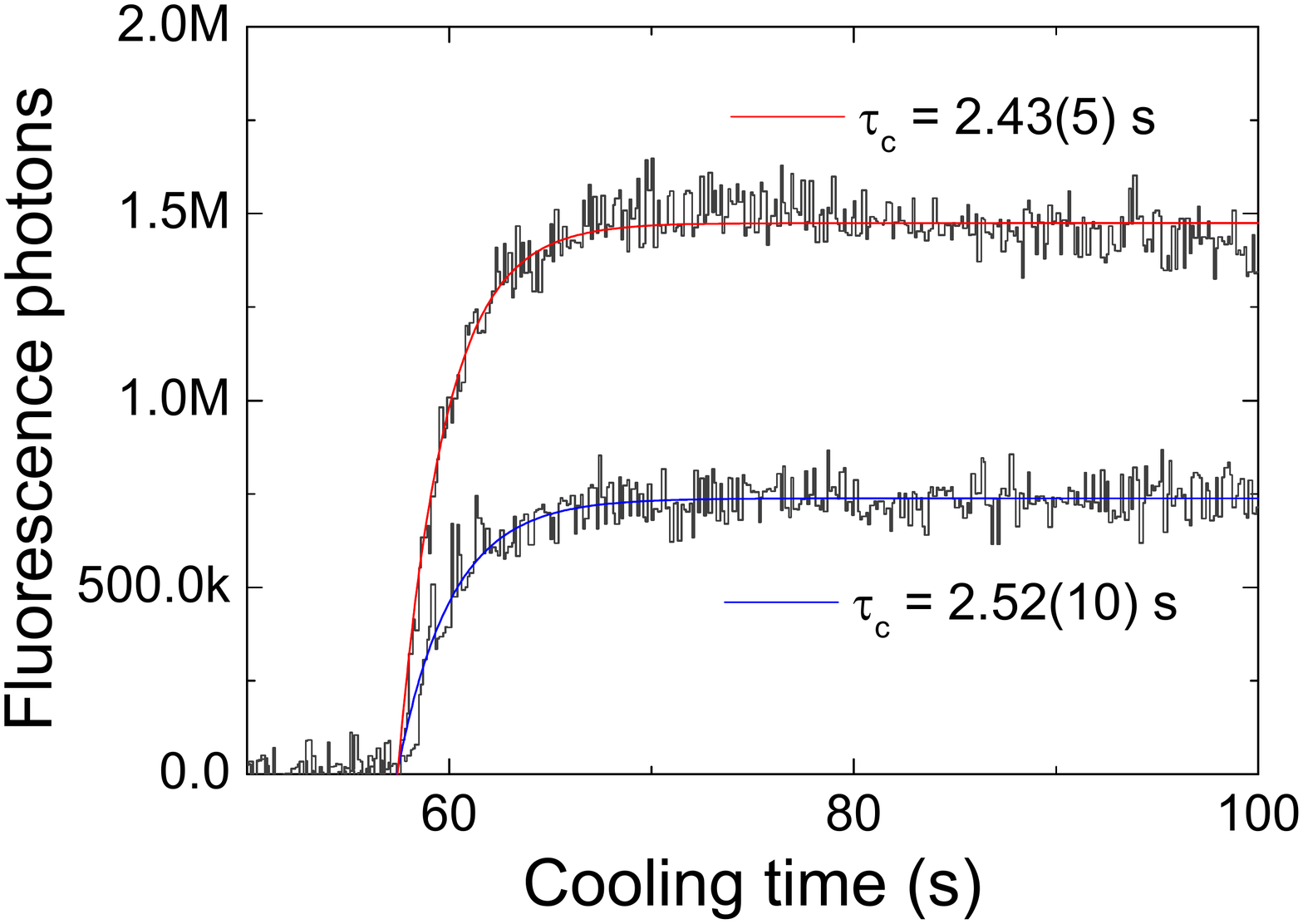}
\vspace{-0.3cm}
\caption{Cooling time for a $^{40}$Ca$^+$ ion cloud when the cooling laser is on resonance (red line), when the laser is detuned 200~MHz below resonance (blue line).\label{fig:4}}
\end{figure}

\noindent where $\Delta \nu (D)$ is the {\it FWHM} of the Gaussian profile originating from the Doppler broadening. For these measurements, the lasers are scanned in similar way as shown in Fig.~\ref{fig:1}. The curve, which is half of the total profile, is fitted with a Voigt function ($V$) as it is shown in Fig.~\ref{fig:new}. Considering the natural linewidth of the Lorentzian profile for $^{40}$Ca$^+$, $\Delta \nu (L)=22$~MHz, following the equation \cite{Oliv1977}

\begin{multline}
\Delta \nu (V) =\\
\frac{1}{2}\left(1.0692\cdot \Delta \nu (L)+\sqrt{0.86639\cdot \Delta ^2 \nu (L)+4\cdot \Delta ^2 \nu (D)}\right), \label{Voigt}
\end{multline}

\noindent and taking $\Delta \nu (V)=339(6)$~MHz from the fit, one obtains $\Delta \nu (D)=327(6)$~MHz, and thus, a temperature of 14.7~K, for the largest ion cloud observed with this optical system. From the number of collected photons, it was possible to estimate or determine the number of ions. For clouds containing around 40 and exactly 12 ions, the ions temperature was 5.6 and 1.5~K, respectively, equivalent to 0.5 and 0.12~meV. These temperatures are smaller than those reported previously of $T=107(7)$~meV \cite{Flec2010}, since only buffer-gas cooling was applied in that case. \\

\noindent Cooling-time measurements were also taken (Fig.~\ref{fig:4}) with the 397-nm lasers on resonance and detuned from resonance by 200~MHz.  The IR lasers were OFF while  loading the trap. The time step for these measurements is 100~ms, which is the shortest exposure time of the EMCCD camera integrated in the acquisition system. The data are fitted using an exponential-grow function. Cooling times are important to utilize this scheme to sympathetically cool ions produced in nuclear reactions, since the applicability to exotic nuclei will be limited by their half-lives \cite{Dela2015}. Further investigations might be carried out on this subject.\\

\section{Laser-cooling of a single ion}

\begin{figure}
\hspace{0cm}
\includegraphics[width=0.5\textwidth]{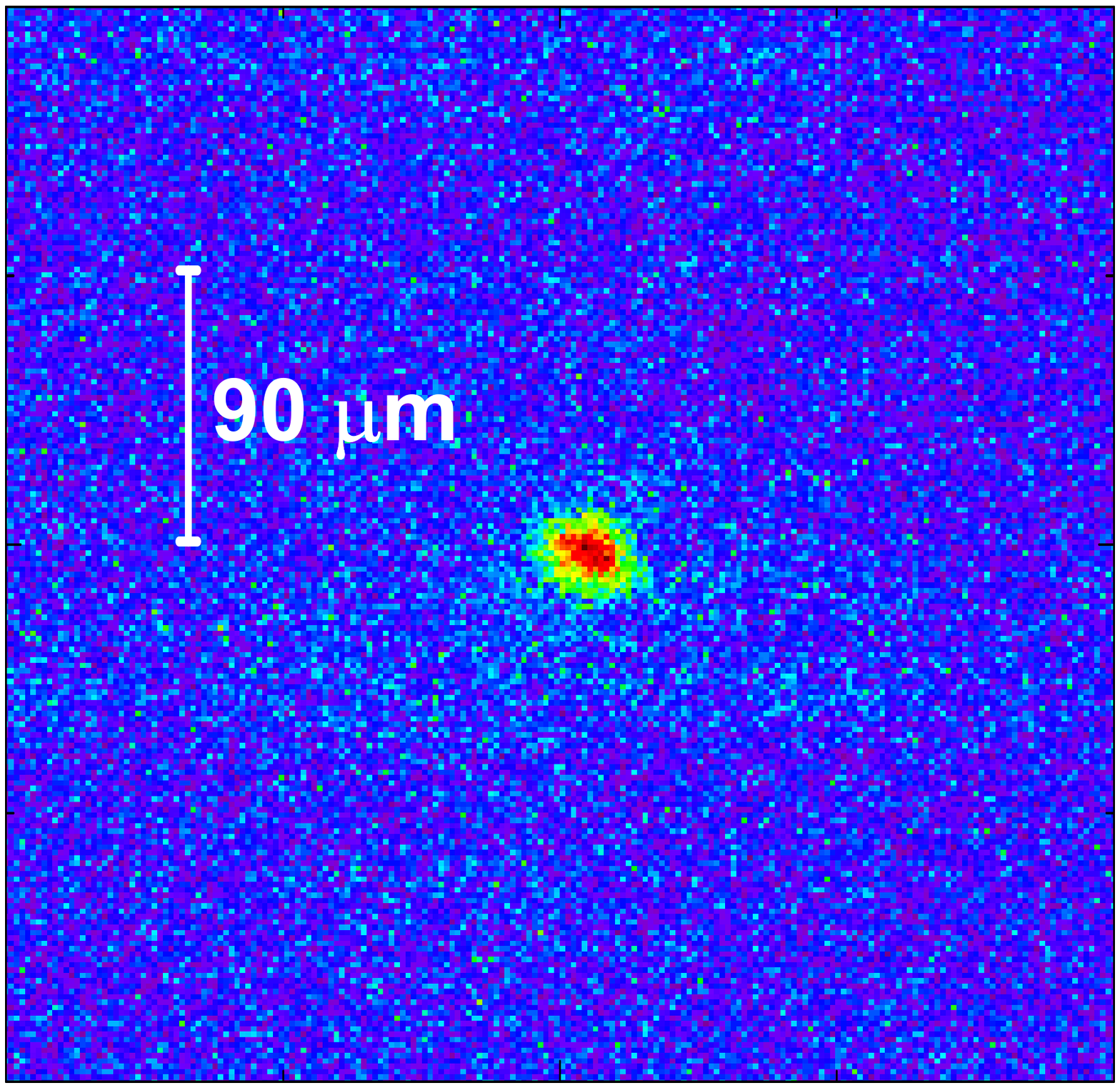}
\vspace{-0.3cm}
\caption{(Color online) Image of a laser-cooled $^{40}$Ca$^+$ ion obtained with the EMCCD camera. The UV- laser frequencies are fixed about 10~MHz below resonance. The exposure time was fixed to 5~seconds and the laser powers of the 397-nm lasers were fixed to 3.8~mW (radial) and 0.9~mW (axial). \label{fig:single}}
\end{figure}

\begin{figure}
\hspace{0cm}
\includegraphics[width=0.5\textwidth]{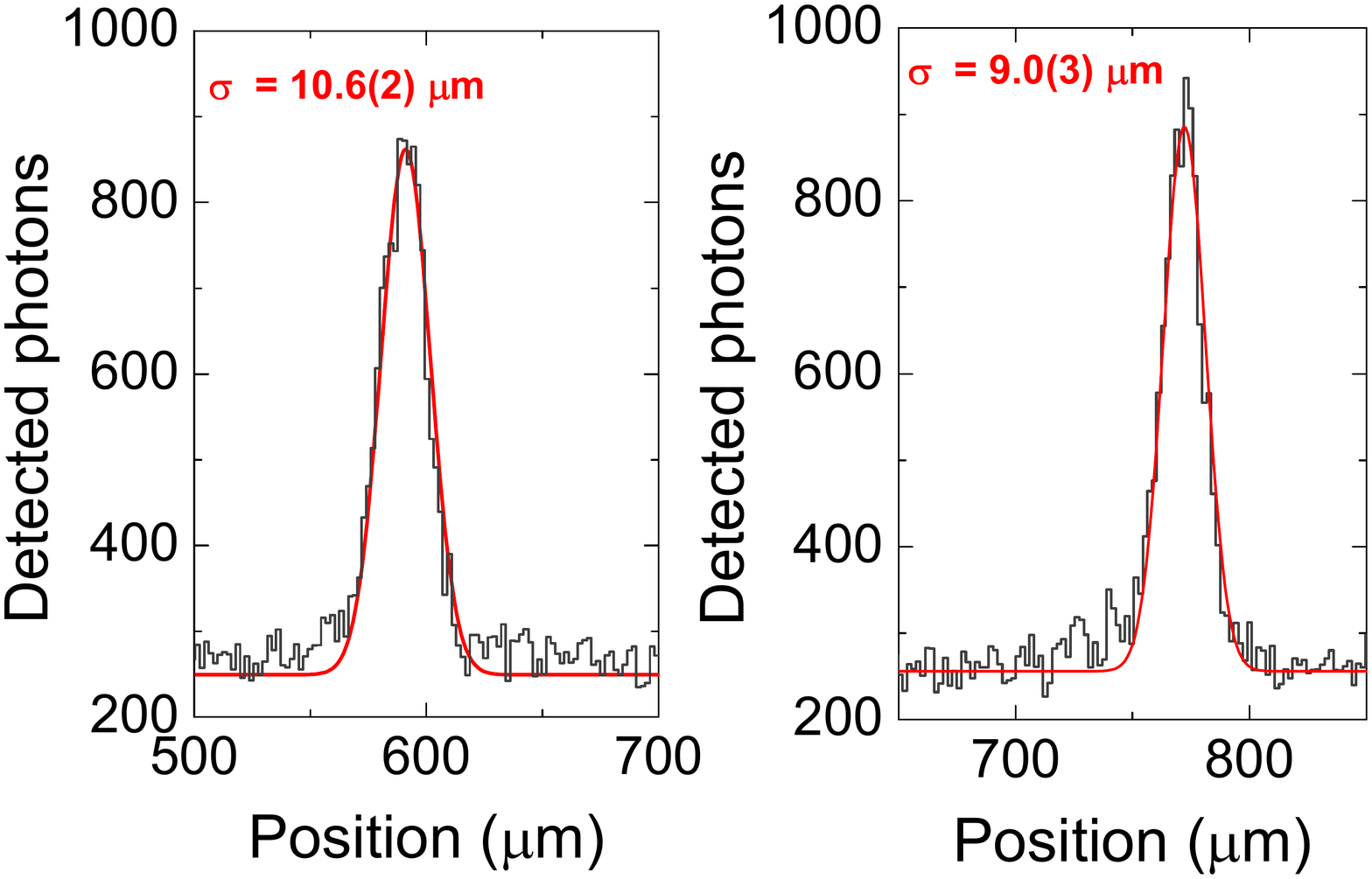}
\vspace{-0.6cm}
\caption{Radial (left panel) and axial (right panel) distributions of the fluorescence photons from a laser-cooled $^{40}$Ca$^+$ ion. These figures are the projections in the radial and axial directions of the signal shown in Fig.~\ref{fig:single}. \label{fig:9_new}}
\end{figure}

In the experiments carried out with this setup, it was possible to start observing differences of one trapped ion in the signal from the fluorescence photons, when the number of stored  ions was around twenty and below. Figure~\ref{fig:single} shows the image of a single $^{40}$Ca$^+$ ion, which has been Doppler-cooled in the open-ring trap, and Fig.~\ref{fig:9_new} shows the projections in the radial and axial planes. The slight difference in the width of the distribution can be assigned to the different potential wells in the axial and in the radial directions, or to subtle distortion effects in the optical system, as inferred from the orientation of the image in the plane of the sensor.  \\

\noindent The laser power for the 397-nm (radial) laser (B2 in Fig.~\ref{fig:1.1}) was varied from 0.304 to 3.8~mW for single-ion measurements. The power for the 397-nm (vertical) laser beam (B1 in Fig.~\ref{fig:1.1}) was fixed. Measurements were taken with and without  the vertical lasers and the outcomes are listed in Tab.~(\ref{power_t}). It is clear that B1 has a small effect on the cooling and just in the axial direction. Taking the beam area and the laser power, the intensity is larger than the saturation intensity for $^{40}$Ca$^+$, of $I_{\hbox{\scriptsize{sat}}}=466$~$\mu $W/mm$^{2}$. However, the ion moves with a very small oscillation amplitude and will only see a small fraction of the laser beam of $\approx 4.7\times 10^{-4}$, and thus the scattering rate might be smaller than expected. \\
 
\begin{table}
\caption{Effect of the B2-laser power on the fluorescence image from a single $^{40}$Ca$^+$ ion. Ratio (axial) and ratio (radial) are the normalized areas resulting from Gaussian fits to the projections of the image in the axial and radial directions, respectively. $\sigma $(axial) are also extracted from the Gaussian fits. $\sigma $(radial) are not listed since the differences are negligible.}\label{power_t}
\begin{ruledtabular}. 
\begin{tabular}{ccccc}
Power (B1)& Power (B2) &Ratio& Ratio& $\sigma$ (axial) \\
(mW)& (mW)& (axial)& (radial)&($\mu$m) \\
\hline
\\
-& 0.304 & 0.13& 0.10& 17.7(5)\\
-& 1.06 & 0.34& 0.29& 14.1(2)\\
-& 1.41 & 0.47& 0.40& 15.8(2)\\
-& 3.8 & 0.89& 0.85& 12.3(1)\\
0.9& 0.304 & 0.21& 0.22& 10.6(3)\\
0.9& 1.06 & 0.40& 0.41& 10.1(2)\\
0.9& 1.41 & 0.48& 0.48& 10.0(2)\\
0.9& 3.8 & 1& 1& 10.1(1)\\
\end{tabular}
\end{ruledtabular}
\end{table}

\noindent  The effect of the RF amplitude on the image is given in Tab.~\ref{shifts}.\\

\begin{table}[b]
\caption{Center of the ion-fluorescence distributions in the radial and axial directions, for different RF voltages. The uncertainty quoted corresponds to 1~$\sigma$, obtained from the Gaussian fit. }\label{shifts}
\begin{ruledtabular}. 
\begin{tabular}{ccccc}
$V_{\scriptsize{\hbox{RF}}}$ (V)& 100& 105.5& 111.5& 117\\
\hline
$q_{\scriptsize{\hbox{z}}}$& 0.368& 0.388& 0.410& 0.430\\
\hline
Radial shift ($\mu $m)& 584(11)& 589(10)& 590(10)& 591(11)\\
\hline
Axial shift ($\mu $m)& 777(8)&  775(8)& 772(9)& 770(9)\\

\end{tabular}
\end{ruledtabular}
\end{table}

\begin{figure}
\hspace{0cm}
\includegraphics[width=0.5\textwidth]{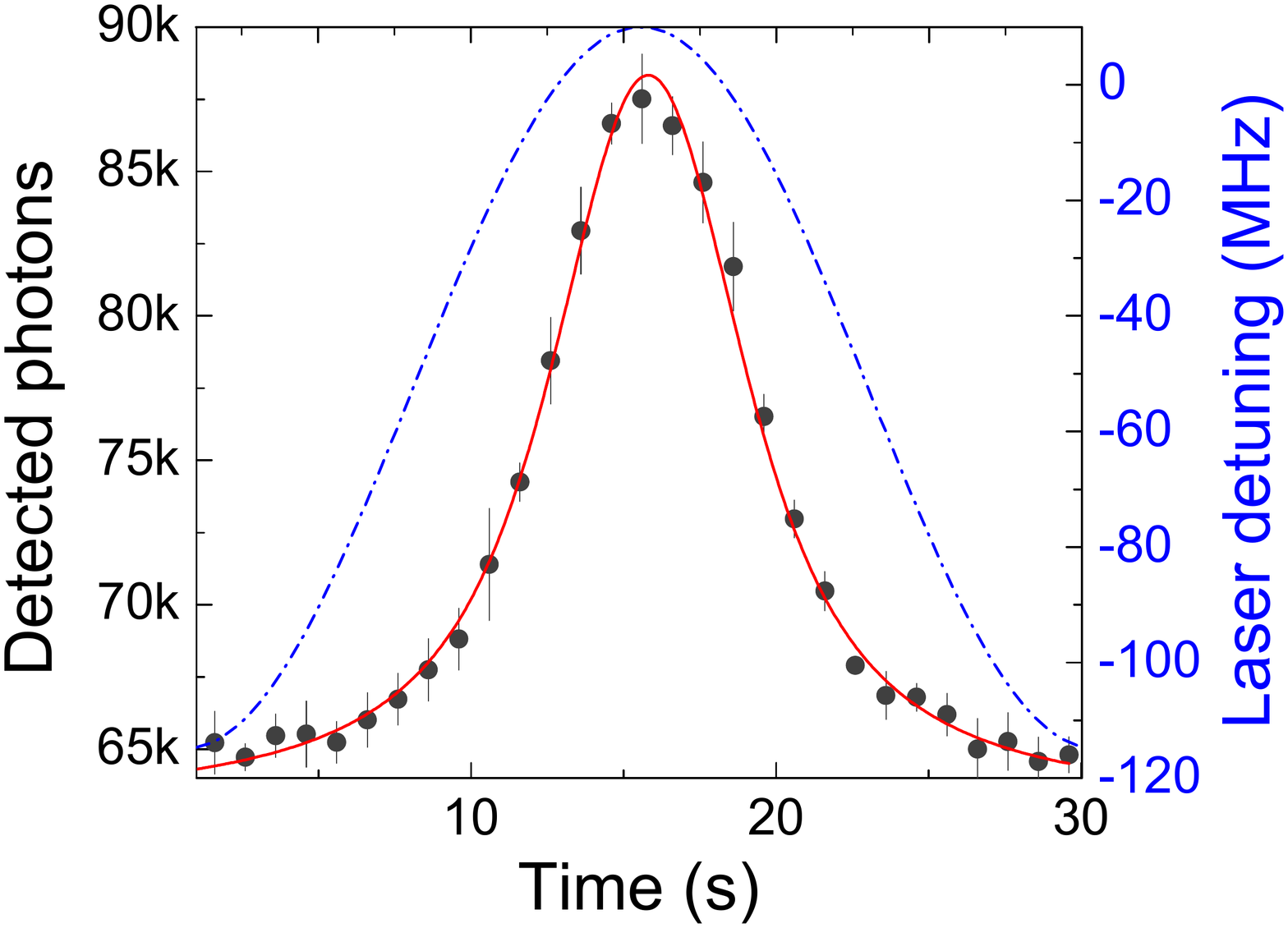}
\vspace{-0.5cm}
\caption{(Color online) Fluorescence photons from a single $^{40}$Ca$^+$ ion, as a function of the B2- laser detuning (blue dotted line). The rest of the laser frequencies were not scanned. The data points represented by filled circles are the average of four measurements. The fit is represented by a red solid line.  \label{fig:8}}
\end{figure}

\begin{figure}
\hspace{0cm}
\includegraphics[width=0.5\textwidth]{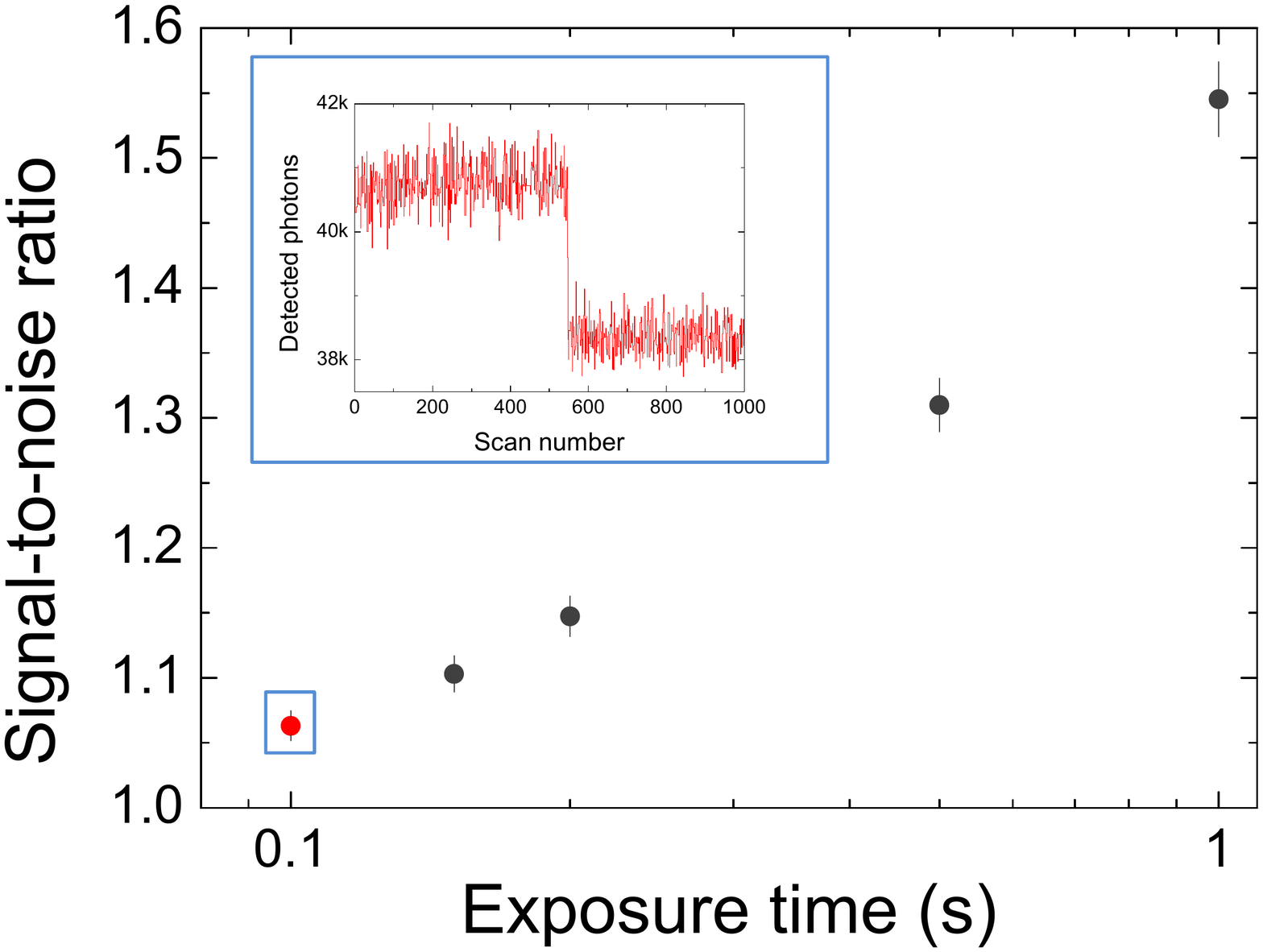}
\vspace{-0.6cm}
\caption{(Color online) Signal-to-noise ratio as a function of the exposure time in the EMCCD camera. The inset shows the signal from one ion and the background signal, when the exposure time is 100~ms.\label{fig:epsart}}
\end{figure}

\noindent An upper limit of the temperature of a single ion, was obtained by scanning the B2-laser, following a sinusoidal function, since this prevents a stronger heating of the ion. The photons collected versus time and the laser frequency detuning are shown in Fig.~\ref{fig:8}. The data are fitted using a Voigt function depending on the time, which is unfolded to frequency using the equation

\begin{equation}
\nu (t) = -52.4 +62.4\cdot  sin\left(\frac{\pi (t-t_c)}{w}\right),
\end{equation}

\noindent where $t_c=8.1$~s, and $w=14.9$~s. $\Delta \nu (L)$ is fixed to $\Delta t $ equivalent to 22~MHz, and the resulting $\Delta t (V)$ yields $\Delta \nu (V)$. Using Eq.~(\ref{Voigt}), one obtains $\Delta \nu =4.23$~MHz, and after substituting in Eq.~(\ref{eq:temperature}), $T_{\scriptsize{\hbox{limit}}}=2.5$~mK, which is in agreement to the expectations. Considering the uncertainty from the fit, $T_{\scriptsize{\hbox{limit}}}^{\scriptsize{\hbox{upper}}}=3.9$~mK,  and $T_{\scriptsize{\hbox{limit}}}^{\scriptsize{\hbox{lower}}}=1.5$~mK. From the temperature of 2.5~mK, it is possible to obtain an average value of the occupational quantum number $< n_z>=<k_BT>/\hbar \omega_z^{sec} \approx 550$ ($\nu _z \approx 94$~kHz). A more precise measurement of the ion temperature would require to probe the quadrupole transition 4s$^2$S$_{1/2}\rightarrow$3d$^2$D$_{5/2}$ at 729~nm, which offers a natural linewidth below 1~Hz.\\

\noindent Most of the measurements presented here have been taken with an exposure time in the EMCCD camera of one second yielding $2.5\times 10^4$~photons per second compared to the expected number from the scattering rate of $\approx \Gamma/4=3.5\times 10^{7}$~s$^{-1}$. This corresponds to an overall efficiency of the optical+detection system of $7\times 10^{-4}$. Shorter acquisition times will provide better timing resolution. Figure~\ref{fig:epsart} shows the signal-to-noise ratio for different exposure times of the EMCCD camera. Times below 100~ms are not possible with the acquisition system currently in use. \\

\begin{figure}
\hspace{0cm}
\includegraphics[width=0.5\textwidth]{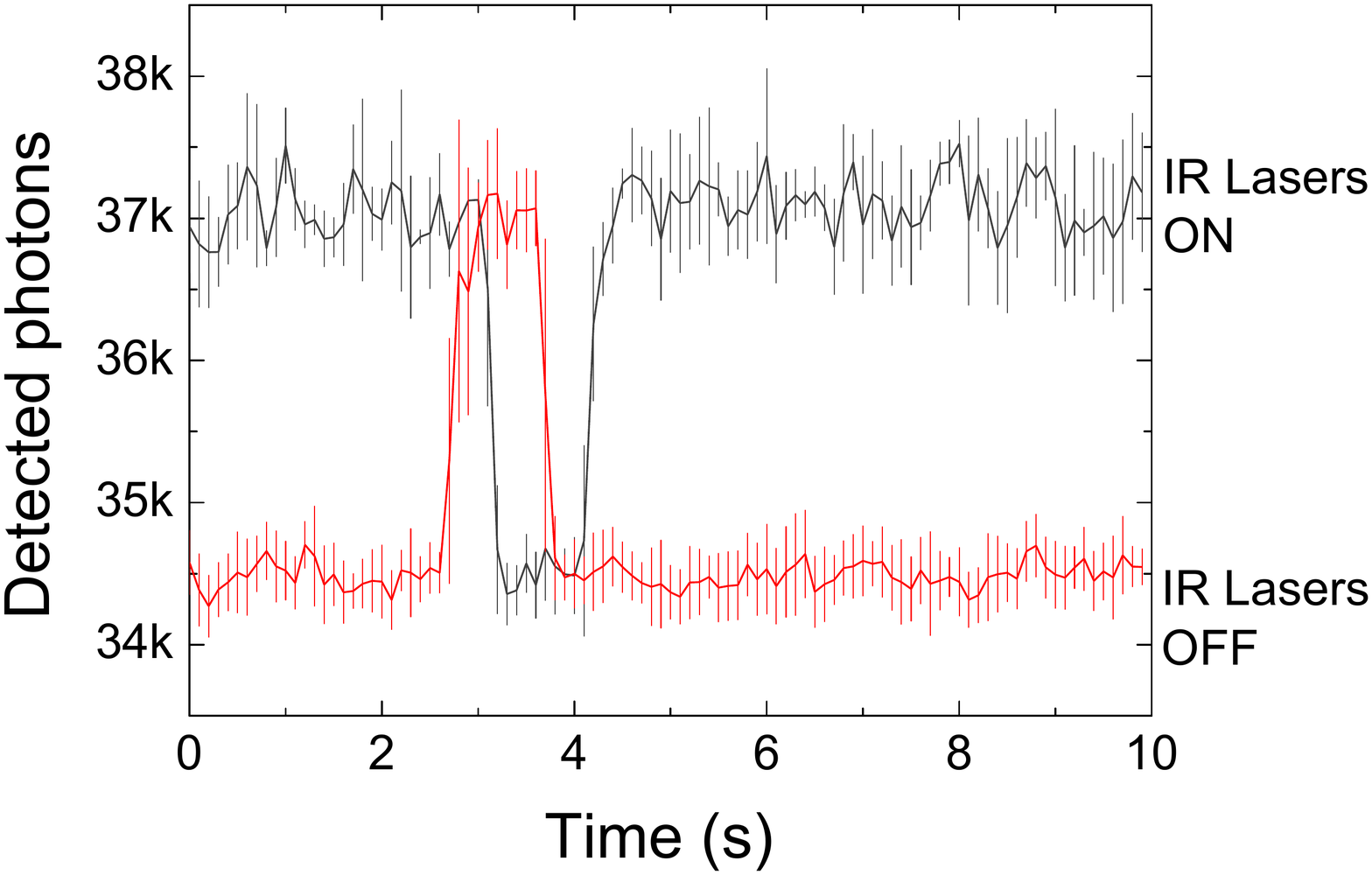}
\vspace{-0.5cm}
\caption{(Color online) Single ion signal using an exposure time of 100~ms in the EMCCD camera. This is the lowest time interval which can be used using the present setup. Two measurements are presented, by stopping the cooling during one second (dark grey) and nine seconds (red). No differences can be observed. \label{fig:epsart2}}
\end{figure}

\noindent First measurements have been taken to probe the heating rate since single-ion sensitivity is important for the final use in the project TRAPSENSOR \cite{Rodr2012}. Figure~\ref{fig:epsart2} shows the fluorescence photons for two different configurations: 1) stopping the cooling for 1~s, and 2) stopping the cooling for 9~s. No heating has been observed, i.e., any energy gained by the ion is below the observable threshold, and therefore, using the maximum laser power, one needs to peform measurements with exposure times below 100~ms.

\section{Conclusions and outlook}

\begin{figure}
\hspace{-0.0cm}
\includegraphics[width=0.5\textwidth]{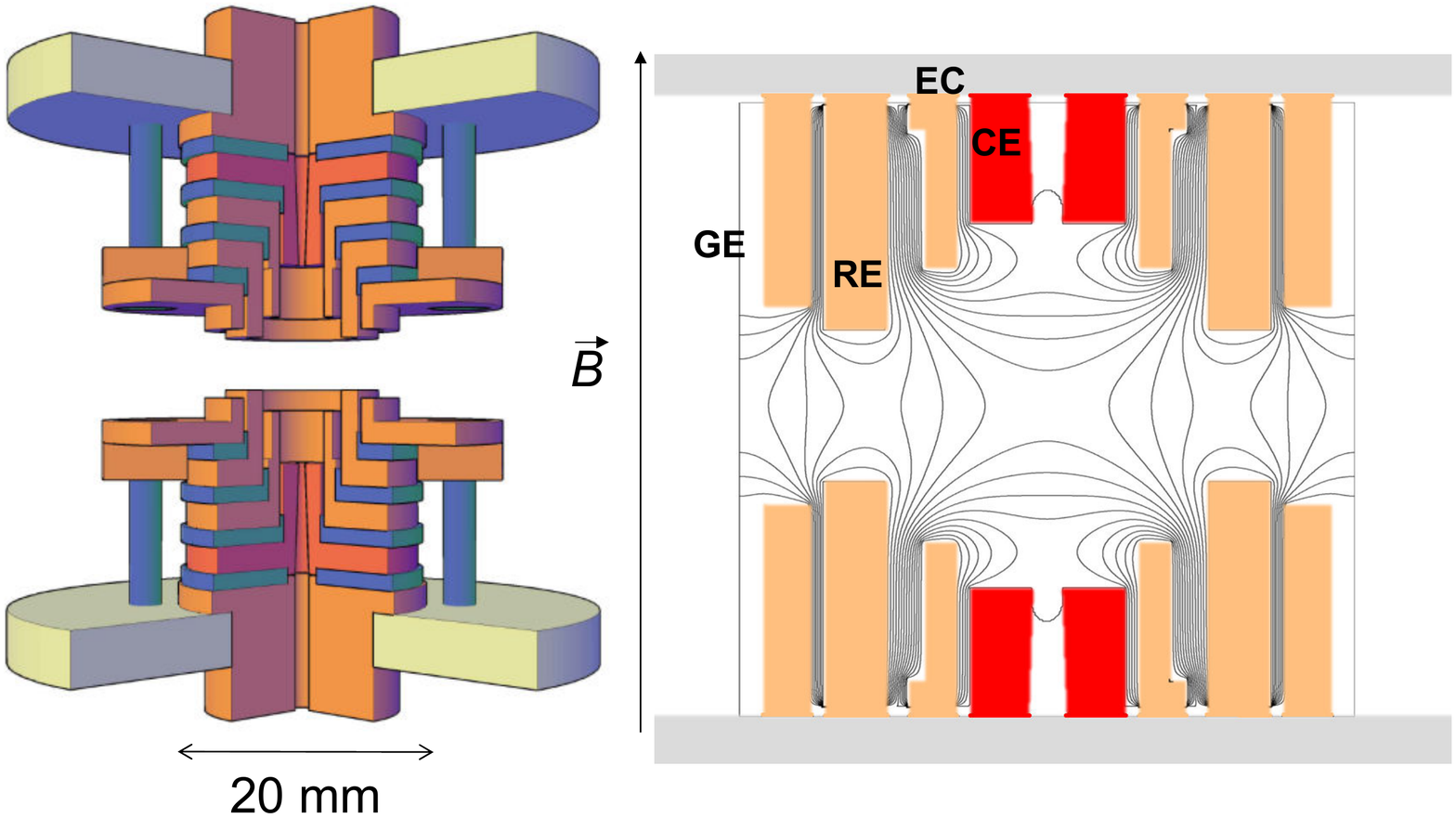}
\vspace{-0.9cm}
\caption{(Color online) Left: Cut from a 3D CAD drawing of a novel Penning trap developed for single-ion trapped mass spectrometry based on optical detection following the design of the trap used for the experiments presented in this paper. Right: Projection of the trap-electrode structure in the axial direction with the equipotential lines. CE stands for coupling electrode, EC for end cap,  RE for ring electrode and GE for ground electrode. For further details see text. \label{fig:trap}}
\end{figure}

Within this work, it has been possible to laser-cool to the Doppler limit, a single $^{40}$Ca$^+$ ion in a large open-ring Paul trap. Besides this feature, this trap has a novel geometry and has served to develop further the Penning-trap system envisaged for the project TRAPSENSOR \cite{Rodr2012}. Figure~\ref{fig:trap} shows a trap, currently installed in a high-homogenous region of a 7-T superconducting solenoid to perform laser-cooling using the laser system shown in Fig.~\ref{fig:1.1} in combination with an electro-optical modulator (EOM). This device is scaled down by a factor of two with respect to the one shown in Figs.~\ref{fig:1.2} and \ref{fig:1.3}, successfully commissioned for this purpose in the course of this work. It is an intermediate step and will be tested shortly at the University of Granada, where it is coupled to a full Penning-trap system comprising a preparation Penning trap, made of a stack of cylinders, a transfer section, and a laser-desorption ion source \cite{Corn2014}. The electrodes colored in orange in the figure, have been modified using numerical methods to add a new electrode (in red), which will be utilized for ion-current detection and later for coupling \cite{Ruiz2014}. 

\begin{acknowledgments}
We acknowledge support from the European Research Council (contract no 278648-TRAPSENSOR) and from the Spanish Ministry of Economy and Competitiveness through the projects FPA2010-14803, FPA2012-32076 and UNGR10-1E-501. We warmly thank G\"unter Werth for fruitful discussions with respect to laser cooling when he was at the University of Granada, and Etienne Li\'enard, and Xaxier Fl\'echard for their comments on the LPCTrap-related part of the paper.

\end{acknowledgments}

\appendix

\nocite{*}
\bibliography{Publication_Daniel_Rodriguez}

\providecommand{\noopsort}[1]{}\providecommand{\singleletter}[1]{#1}%
\begin{thebibliography}{28}%
\makeatletter
\providecommand \@ifxundefined [1]{%
 \@ifx{#1\undefined}
}%
\providecommand \@ifnum [1]{%
 \ifnum #1\expandafter \@firstoftwo
 \else \expandafter \@secondoftwo
 \fi
}%
\providecommand \@ifx [1]{%
 \ifx #1\expandafter \@firstoftwo
 \else \expandafter \@secondoftwo
 \fi
}%
\providecommand \natexlab [1]{#1}%
\providecommand \enquote  [1]{``#1''}%
\providecommand \bibnamefont  [1]{#1}%
\providecommand \bibfnamefont [1]{#1}%
\providecommand \citenamefont [1]{#1}%
\providecommand \href@noop [0]{\@secondoftwo}%
\providecommand \href [0]{\begingroup \@sanitize@url \@href}%
\providecommand \@href[1]{\@@startlink{#1}\@@href}%
\providecommand \@@href[1]{\endgroup#1\@@endlink}%
\providecommand \@sanitize@url [0]{\catcode `\\12\catcode `\$12\catcode
  `\&12\catcode `\#12\catcode `\^12\catcode `\_12\catcode `\%12\relax}%
\providecommand \@@startlink[1]{}%
\providecommand \@@endlink[0]{}%
\providecommand \url  [0]{\begingroup\@sanitize@url \@url }%
\providecommand \@url [1]{\endgroup\@href {#1}{\urlprefix }}%
\providecommand \urlprefix  [0]{URL }%
\providecommand \Eprint [0]{\href }%
\providecommand \doibase [0]{http://dx.doi.org/}%
\providecommand \selectlanguage [0]{\@gobble}%
\providecommand \bibinfo  [0]{\@secondoftwo}%
\providecommand \bibfield  [0]{\@secondoftwo}%
\providecommand \translation [1]{[#1]}%
\providecommand \BibitemOpen [0]{}%
\providecommand \bibitemStop [0]{}%
\providecommand \bibitemNoStop [0]{.\EOS\space}%
\providecommand \EOS [0]{\spacefactor3000\relax}%
\providecommand \BibitemShut  [1]{\csname bibitem#1\endcsname}%
\let\auto@bib@innerbib\@empty
\bibitem [{\citenamefont {Leibfried}\ \emph {et~al.}(2003)\citenamefont
  {Leibfried}, \citenamefont {Blatt}, \citenamefont {Monroe},\ and\
  \citenamefont {Wineland}}]{Leib2003}%
  \BibitemOpen
  \bibfield  {author} {\bibinfo {author} {\bibfnamefont {D.}~\bibnamefont
  {Leibfried}}, \bibinfo {author} {\bibfnamefont {R.}~\bibnamefont {Blatt}},
  \bibinfo {author} {\bibfnamefont {C.}~\bibnamefont {Monroe}}, \ and\ \bibinfo
  {author} {\bibfnamefont {D.}~\bibnamefont {Wineland}},\ }\href@noop {}
  {\bibfield  {journal} {\bibinfo  {journal} {Rev. Mod. Phys.}\ }\textbf
  {\bibinfo {volume} {75}},\ \bibinfo {pages} {281} (\bibinfo {year}
  {2003})}\BibitemShut {NoStop}%
\bibitem [{\citenamefont {Chwalla}\ \emph {et~al.}(2009)\citenamefont
  {Chwalla}, \citenamefont {Benhelm}, \citenamefont {Kim}, \citenamefont
  {Kirchmair}, \citenamefont {Monz}, \citenamefont {Riebe}, \citenamefont
  {Schindler}, \citenamefont {Villar}, \citenamefont {H\"ansel}, \citenamefont
  {Roos}, \citenamefont {Blatt}, \citenamefont {Abgrall}, \citenamefont
  {Santarelli}, \citenamefont {Rovera},\ and\ \citenamefont
  {Laurent}}]{Chwa2009}%
  \BibitemOpen
  \bibfield  {author} {\bibinfo {author} {\bibfnamefont {M.}~\bibnamefont
  {Chwalla}}, \bibinfo {author} {\bibfnamefont {J.}~\bibnamefont {Benhelm}},
  \bibinfo {author} {\bibfnamefont {K.}~\bibnamefont {Kim}}, \bibinfo {author}
  {\bibfnamefont {G.}~\bibnamefont {Kirchmair}}, \bibinfo {author}
  {\bibfnamefont {T.}~\bibnamefont {Monz}}, \bibinfo {author} {\bibfnamefont
  {M.}~\bibnamefont {Riebe}}, \bibinfo {author} {\bibfnamefont
  {P.}~\bibnamefont {Schindler}}, \bibinfo {author} {\bibfnamefont
  {A.}~\bibnamefont {Villar}}, \bibinfo {author} {\bibfnamefont
  {W.}~\bibnamefont {H\"ansel}}, \bibinfo {author} {\bibfnamefont
  {C.}~\bibnamefont {Roos}}, \bibinfo {author} {\bibfnamefont {R.}~\bibnamefont
  {Blatt}}, \bibinfo {author} {\bibfnamefont {M.}~\bibnamefont {Abgrall}},
  \bibinfo {author} {\bibfnamefont {G.}~\bibnamefont {Santarelli}}, \bibinfo
  {author} {\bibfnamefont {G.}~\bibnamefont {Rovera}}, \ and\ \bibinfo {author}
  {\bibfnamefont {P.}~\bibnamefont {Laurent}},\ }\href@noop {} {\bibfield
  {journal} {\bibinfo  {journal} {Phys. Rev. Lett.}\ }\textbf {\bibinfo
  {volume} {102}},\ \bibinfo {pages} {023002} (\bibinfo {year}
  {2009})}\BibitemShut {NoStop}%
\bibitem [{\citenamefont {Schmidt}\ \emph {et~al.}(2005)\citenamefont
  {Schmidt}, \citenamefont {Rosenband}, \citenamefont {Langer}, \citenamefont
  {Itano}, \citenamefont {Bergquist},\ and\ \citenamefont
  {Wineland}}]{Schm2005}%
  \BibitemOpen
  \bibfield  {author} {\bibinfo {author} {\bibfnamefont {P.~O.}\ \bibnamefont
  {Schmidt}}, \bibinfo {author} {\bibfnamefont {T.}~\bibnamefont {Rosenband}},
  \bibinfo {author} {\bibfnamefont {C.}~\bibnamefont {Langer}}, \bibinfo
  {author} {\bibfnamefont {W.~M.}\ \bibnamefont {Itano}}, \bibinfo {author}
  {\bibfnamefont {J.~C.}\ \bibnamefont {Bergquist}}, \ and\ \bibinfo {author}
  {\bibfnamefont {D.~J.}\ \bibnamefont {Wineland}},\ }\href@noop {} {\bibfield
  {journal} {\bibinfo  {journal} {Science}\ }\textbf {\bibinfo {volume}
  {309}},\ \bibinfo {pages} {749} (\bibinfo {year} {2005})}\BibitemShut
  {NoStop}%
\bibitem [{\citenamefont {Blatt}\ and\ \citenamefont {Roos}(2012)}]{blat2012}%
  \BibitemOpen
  \bibfield  {author} {\bibinfo {author} {\bibfnamefont {R.}~\bibnamefont
  {Blatt}}\ and\ \bibinfo {author} {\bibfnamefont {C.~F.}\ \bibnamefont
  {Roos}},\ }\href@noop {} {\bibfield  {journal} {\bibinfo  {journal} {Nature
  Phys.}\ }\textbf {\bibinfo {volume} {8}},\ \bibinfo {pages} {277} (\bibinfo
  {year} {2012})}\BibitemShut {NoStop}%
\bibitem [{\citenamefont {Kielpinski}, \citenamefont {Monroe},\ and\
  \citenamefont {Wineland}(2002)}]{Kiel2002}%
  \BibitemOpen
  \bibfield  {author} {\bibinfo {author} {\bibfnamefont {D.}~\bibnamefont
  {Kielpinski}}, \bibinfo {author} {\bibfnamefont {C.}~\bibnamefont {Monroe}},
  \ and\ \bibinfo {author} {\bibfnamefont {D.~J.}\ \bibnamefont {Wineland}},\
  }\href@noop {} {\bibfield  {journal} {\bibinfo  {journal} {Nature}\ }\textbf
  {\bibinfo {volume} {417}},\ \bibinfo {pages} {709} (\bibinfo {year}
  {2002})}\BibitemShut {NoStop}%
\bibitem [{\citenamefont {Mitchell}\ \emph {et~al.}(1998)\citenamefont
  {Mitchell}, \citenamefont {Bollinger}, \citenamefont {Dubin}, \citenamefont
  {Huang}, \citenamefont {Itano},\ and\ \citenamefont {Baughman}}]{Mitc1998}%
  \BibitemOpen
  \bibfield  {author} {\bibinfo {author} {\bibfnamefont {T.~B.}\ \bibnamefont
  {Mitchell}}, \bibinfo {author} {\bibfnamefont {J.~J.}\ \bibnamefont
  {Bollinger}}, \bibinfo {author} {\bibfnamefont {D.~H.~E.}\ \bibnamefont
  {Dubin}}, \bibinfo {author} {\bibfnamefont {X.~P.}\ \bibnamefont {Huang}},
  \bibinfo {author} {\bibfnamefont {W.~M.}\ \bibnamefont {Itano}}, \ and\
  \bibinfo {author} {\bibfnamefont {R.~H.}\ \bibnamefont {Baughman}},\
  }\href@noop {} {\bibfield  {journal} {\bibinfo  {journal} {Science}\ }\textbf
  {\bibinfo {volume} {282}},\ \bibinfo {pages} {1290} (\bibinfo {year}
  {1998})}\BibitemShut {NoStop}%
\bibitem [{\citenamefont {Madavia}\ \emph {et~al.}(2013)\citenamefont
  {Madavia}, \citenamefont {Goddwin}, \citenamefont {Stutter}, \citenamefont
  {Bharadia}, \citenamefont {Segal},\ and\ \citenamefont
  {Thompson}}]{Mada2013}%
  \BibitemOpen
  \bibfield  {author} {\bibinfo {author} {\bibfnamefont {S.}~\bibnamefont
  {Madavia}}, \bibinfo {author} {\bibfnamefont {J.~F.}\ \bibnamefont
  {Goddwin}}, \bibinfo {author} {\bibfnamefont {G.}~\bibnamefont {Stutter}},
  \bibinfo {author} {\bibfnamefont {S.}~\bibnamefont {Bharadia}}, \bibinfo
  {author} {\bibfnamefont {D.~M.}\ \bibnamefont {Segal}}, \ and\ \bibinfo
  {author} {\bibfnamefont {R.~C.}\ \bibnamefont {Thompson}},\ }\href@noop {}
  {\bibfield  {journal} {\bibinfo  {journal} {Nature Communications}\ }\textbf
  {\bibinfo {volume} {4}},\ \bibinfo {pages} {2571} (\bibinfo {year}
  {2013})}\BibitemShut {NoStop}%
\bibitem [{\citenamefont {Andelkovic}\ \emph {et~al.}(2013)\citenamefont
  {Andelkovic}, \citenamefont {Cazan}, \citenamefont {N\"ortersh\"auser},
  \citenamefont {Bharadia}, \citenamefont {Segal}, \citenamefont {Thompson},
  \citenamefont {Johren}, \citenamefont {Vollbrecht}, \citenamefont {Hannen},\
  and\ \citenamefont {Vogel}}]{Ande2013}%
  \BibitemOpen
  \bibfield  {author} {\bibinfo {author} {\bibfnamefont {Z.}~\bibnamefont
  {Andelkovic}}, \bibinfo {author} {\bibfnamefont {R.}~\bibnamefont {Cazan}},
  \bibinfo {author} {\bibfnamefont {W.}~\bibnamefont {N\"ortersh\"auser}},
  \bibinfo {author} {\bibfnamefont {S.}~\bibnamefont {Bharadia}}, \bibinfo
  {author} {\bibfnamefont {D.~M.}\ \bibnamefont {Segal}}, \bibinfo {author}
  {\bibfnamefont {R.~C.}\ \bibnamefont {Thompson}}, \bibinfo {author}
  {\bibfnamefont {R.}~\bibnamefont {Johren}}, \bibinfo {author} {\bibfnamefont
  {J.}~\bibnamefont {Vollbrecht}}, \bibinfo {author} {\bibfnamefont
  {V.}~\bibnamefont {Hannen}}, \ and\ \bibinfo {author} {\bibfnamefont
  {M.}~\bibnamefont {Vogel}},\ }\href@noop {} {\bibfield  {journal} {\bibinfo
  {journal} {Phys. Rev. A}\ }\textbf {\bibinfo {volume} {87}},\ \bibinfo
  {pages} {033423} (\bibinfo {year} {2013})}\BibitemShut {NoStop}%
\bibitem [{\citenamefont {Blaum}, \citenamefont {Dilling},\ and\ \citenamefont
  {N\"ortersh\"auser}(2013)}]{Blau2013}%
  \BibitemOpen
  \bibfield  {author} {\bibinfo {author} {\bibfnamefont {K.}~\bibnamefont
  {Blaum}}, \bibinfo {author} {\bibfnamefont {J.}~\bibnamefont {Dilling}}, \
  and\ \bibinfo {author} {\bibfnamefont {W.}~\bibnamefont
  {N\"ortersh\"auser}},\ }\href@noop {} {\bibfield  {journal} {\bibinfo
  {journal} {Phys. Scrip.}\ }\textbf {\bibinfo {volume} {T152}},\ \bibinfo
  {pages} {014017} (\bibinfo {year} {2013})}\BibitemShut {NoStop}%
\bibitem [{\citenamefont {Rodr\'iguez}(2012)}]{Rodr2012}%
  \BibitemOpen
  \bibfield  {author} {\bibinfo {author} {\bibfnamefont {D.}~\bibnamefont
  {Rodr\'iguez}},\ }\href@noop {} {\bibfield  {journal} {\bibinfo  {journal}
  {Appl. Phys. B: Lasers O.}\ }\textbf {\bibinfo {volume} {107}},\ \bibinfo
  {pages} {1031} (\bibinfo {year} {2012})}\BibitemShut {NoStop}%
\bibitem [{\citenamefont {Redshaw}, \citenamefont {McDaniel},\ and\
  \citenamefont {Myers}(2008)}]{Reds2008}%
  \BibitemOpen
  \bibfield  {author} {\bibinfo {author} {\bibfnamefont {M.}~\bibnamefont
  {Redshaw}}, \bibinfo {author} {\bibfnamefont {J.~M.}\ \bibnamefont
  {McDaniel}}, \ and\ \bibinfo {author} {\bibfnamefont {E.~G.}\ \bibnamefont
  {Myers}},\ }\href@noop {} {\bibfield  {journal} {\bibinfo  {journal} {Phys.
  Rev. Lett.}\ }\textbf {\bibinfo {volume} {100}},\ \bibinfo {pages} {093002}
  (\bibinfo {year} {2008})}\BibitemShut {NoStop}%
\bibitem [{\citenamefont {Block}\ \emph {et~al.}(2005)\citenamefont {Block},
  \citenamefont {Ackermann}, \citenamefont {Blaum}, \citenamefont {Droese},
  \citenamefont {Dworschak}, \citenamefont {Eliseev}, \citenamefont
  {Fleckenstein}, \citenamefont {Haettner}, \citenamefont {Herfurth},
  \citenamefont {Heßberger}, \citenamefont {Hofmann}, \citenamefont
  {Ketelaer}, \citenamefont {Ketter}, \citenamefont {Kluge}, \citenamefont
  {Marx}, \citenamefont {Mazzocco}, \citenamefont {Novikov}, \citenamefont
  {Plass}, \citenamefont {Popeko}, \citenamefont {Rahaman}, \citenamefont
  {Rodr\'iguez}, \citenamefont {Scheidenberger}, \citenamefont {Schweikhard},
  \citenamefont {Thirolf}, \citenamefont {Vorobyev},\ and\ \citenamefont
  {Weber}}]{Bloc2005}%
  \BibitemOpen
  \bibfield  {author} {\bibinfo {author} {\bibfnamefont {M.}~\bibnamefont
  {Block}}, \bibinfo {author} {\bibfnamefont {D.}~\bibnamefont {Ackermann}},
  \bibinfo {author} {\bibfnamefont {K.}~\bibnamefont {Blaum}}, \bibinfo
  {author} {\bibfnamefont {C.}~\bibnamefont {Droese}}, \bibinfo {author}
  {\bibfnamefont {M.}~\bibnamefont {Dworschak}}, \bibinfo {author}
  {\bibfnamefont {S.}~\bibnamefont {Eliseev}}, \bibinfo {author} {\bibfnamefont
  {T.}~\bibnamefont {Fleckenstein}}, \bibinfo {author} {\bibfnamefont
  {E.}~\bibnamefont {Haettner}}, \bibinfo {author} {\bibfnamefont
  {F.}~\bibnamefont {Herfurth}}, \bibinfo {author} {\bibfnamefont {F.~P.}\
  \bibnamefont {Heßberger}}, \bibinfo {author} {\bibfnamefont
  {S.}~\bibnamefont {Hofmann}}, \bibinfo {author} {\bibfnamefont
  {J.}~\bibnamefont {Ketelaer}}, \bibinfo {author} {\bibfnamefont
  {J.}~\bibnamefont {Ketter}}, \bibinfo {author} {\bibfnamefont {H.-J.}\
  \bibnamefont {Kluge}}, \bibinfo {author} {\bibfnamefont {G.}~\bibnamefont
  {Marx}}, \bibinfo {author} {\bibfnamefont {M.}~\bibnamefont {Mazzocco}},
  \bibinfo {author} {\bibfnamefont {Y.~N.}\ \bibnamefont {Novikov}}, \bibinfo
  {author} {\bibfnamefont {W.~R.}\ \bibnamefont {Plass}}, \bibinfo {author}
  {\bibfnamefont {A.}~\bibnamefont {Popeko}}, \bibinfo {author} {\bibfnamefont
  {S.}~\bibnamefont {Rahaman}}, \bibinfo {author} {\bibfnamefont
  {D.}~\bibnamefont {Rodr\'iguez}}, \bibinfo {author} {\bibfnamefont
  {C.}~\bibnamefont {Scheidenberger}}, \bibinfo {author} {\bibfnamefont
  {L.}~\bibnamefont {Schweikhard}}, \bibinfo {author} {\bibfnamefont {P.~G.}\
  \bibnamefont {Thirolf}}, \bibinfo {author} {\bibfnamefont {G.~K.}\
  \bibnamefont {Vorobyev}}, \ and\ \bibinfo {author} {\bibfnamefont
  {C.}~\bibnamefont {Weber}},\ }\href@noop {} {\bibfield  {journal} {\bibinfo
  {journal} {Eur. Phys. J. A}\ }\textbf {\bibinfo {volume} {25}},\ \bibinfo
  {pages} {49} (\bibinfo {year} {2005})}\BibitemShut {NoStop}%
\bibitem [{\citenamefont {Block}\ \emph {et~al.}(2010)\citenamefont {Block},
  \citenamefont {Ackermann}, \citenamefont {Blaum}, \citenamefont {Droese},
  \citenamefont {Dworschak}, \citenamefont {Eliseev}, \citenamefont
  {Fleckenstein}, \citenamefont {Haettner}, \citenamefont {Herfurth},
  \citenamefont {Heßberger}, \citenamefont {Hofmann}, \citenamefont
  {Ketelaer}, \citenamefont {Ketter}, \citenamefont {Kluge}, \citenamefont
  {Marx}, \citenamefont {Mazzocco}, \citenamefont {Novikov}, \citenamefont
  {Plass}, \citenamefont {Popeko}, \citenamefont {Rahaman}, \citenamefont
  {Rodr\'iguez}, \citenamefont {Scheidenberger}, \citenamefont {Schweikhard},
  \citenamefont {Thirolf}, \citenamefont {Vorobyev},\ and\ \citenamefont
  {Weber}}]{Bloc2010}%
  \BibitemOpen
  \bibfield  {author} {\bibinfo {author} {\bibfnamefont {M.}~\bibnamefont
  {Block}}, \bibinfo {author} {\bibfnamefont {D.}~\bibnamefont {Ackermann}},
  \bibinfo {author} {\bibfnamefont {K.}~\bibnamefont {Blaum}}, \bibinfo
  {author} {\bibfnamefont {C.}~\bibnamefont {Droese}}, \bibinfo {author}
  {\bibfnamefont {M.}~\bibnamefont {Dworschak}}, \bibinfo {author}
  {\bibfnamefont {S.}~\bibnamefont {Eliseev}}, \bibinfo {author} {\bibfnamefont
  {T.}~\bibnamefont {Fleckenstein}}, \bibinfo {author} {\bibfnamefont
  {E.}~\bibnamefont {Haettner}}, \bibinfo {author} {\bibfnamefont
  {F.}~\bibnamefont {Herfurth}}, \bibinfo {author} {\bibfnamefont {F.~P.}\
  \bibnamefont {Heßberger}}, \bibinfo {author} {\bibfnamefont
  {S.}~\bibnamefont {Hofmann}}, \bibinfo {author} {\bibfnamefont
  {J.}~\bibnamefont {Ketelaer}}, \bibinfo {author} {\bibfnamefont
  {J.}~\bibnamefont {Ketter}}, \bibinfo {author} {\bibfnamefont {H.-J.}\
  \bibnamefont {Kluge}}, \bibinfo {author} {\bibfnamefont {G.}~\bibnamefont
  {Marx}}, \bibinfo {author} {\bibfnamefont {M.}~\bibnamefont {Mazzocco}},
  \bibinfo {author} {\bibfnamefont {Y.~N.}\ \bibnamefont {Novikov}}, \bibinfo
  {author} {\bibfnamefont {W.~R.}\ \bibnamefont {Plass}}, \bibinfo {author}
  {\bibfnamefont {A.}~\bibnamefont {Popeko}}, \bibinfo {author} {\bibfnamefont
  {S.}~\bibnamefont {Rahaman}}, \bibinfo {author} {\bibfnamefont
  {D.}~\bibnamefont {Rodr\'iguez}}, \bibinfo {author} {\bibfnamefont
  {C.}~\bibnamefont {Scheidenberger}}, \bibinfo {author} {\bibfnamefont
  {L.}~\bibnamefont {Schweikhard}}, \bibinfo {author} {\bibfnamefont {P.~G.}\
  \bibnamefont {Thirolf}}, \bibinfo {author} {\bibfnamefont {G.~K.}\
  \bibnamefont {Vorobyev}}, \ and\ \bibinfo {author} {\bibfnamefont
  {C.}~\bibnamefont {Weber}},\ }\href@noop {} {\bibfield  {journal} {\bibinfo
  {journal} {Nature}\ }\textbf {\bibinfo {volume} {463}},\ \bibinfo {pages}
  {785} (\bibinfo {year} {2010})}\BibitemShut {NoStop}%
\bibitem [{\citenamefont {Heinzen}\ and\ \citenamefont
  {Weinland}(1990)}]{Wine1990}%
  \BibitemOpen
  \bibfield  {author} {\bibinfo {author} {\bibfnamefont {D.~J.}\ \bibnamefont
  {Heinzen}}\ and\ \bibinfo {author} {\bibfnamefont {D.~J.}\ \bibnamefont
  {Weinland}},\ }\href@noop {} {\bibfield  {journal} {\bibinfo  {journal}
  {Phys. Rev. A}\ }\textbf {\bibinfo {volume} {42(5)}},\ \bibinfo {pages}
  {2977} (\bibinfo {year} {1990})}\BibitemShut {NoStop}%
\bibitem [{\citenamefont {Brown}\ \emph {et~al.}(2011)\citenamefont {Brown},
  \citenamefont {Ospelkaus}, \citenamefont {Colombe}, \citenamefont {Wilson},
  \citenamefont {Leibfried},\ and\ \citenamefont {Wineland}}]{Brow2011}%
  \BibitemOpen
  \bibfield  {author} {\bibinfo {author} {\bibfnamefont {K.~R.}\ \bibnamefont
  {Brown}}, \bibinfo {author} {\bibfnamefont {C.}~\bibnamefont {Ospelkaus}},
  \bibinfo {author} {\bibfnamefont {Y.}~\bibnamefont {Colombe}}, \bibinfo
  {author} {\bibfnamefont {A.~C.}\ \bibnamefont {Wilson}}, \bibinfo {author}
  {\bibfnamefont {D.}~\bibnamefont {Leibfried}}, \ and\ \bibinfo {author}
  {\bibfnamefont {D.~J.}\ \bibnamefont {Wineland}},\ }\href@noop {} {\bibfield
  {journal} {\bibinfo  {journal} {Nature}\ }\textbf {\bibinfo {volume} {471}},\
  \bibinfo {pages} {196} (\bibinfo {year} {2011})}\BibitemShut {NoStop}%
\bibitem [{\citenamefont {Rodr\'iguez}\ \emph {et~al.}(2006)\citenamefont
  {Rodr\'iguez}, \citenamefont {M\'ery}, \citenamefont {Ban}, \citenamefont
  {Br\'egeault}, \citenamefont {Darius}, \citenamefont {Durand}, \citenamefont
  {Fl\'echard}, \citenamefont {Herbane}, \citenamefont {Labalme}, \citenamefont
  {Li\'enard}, \citenamefont {Mauger}, \citenamefont {Merrer}, \citenamefont
  {Naviliat-Cuncic}, \citenamefont {Thomas},\ and\ \citenamefont
  {Vandamme}}]{Rodr2006}%
  \BibitemOpen
  \bibfield  {author} {\bibinfo {author} {\bibfnamefont {D.}~\bibnamefont
  {Rodr\'iguez}}, \bibinfo {author} {\bibfnamefont {A.}~\bibnamefont {M\'ery}},
  \bibinfo {author} {\bibfnamefont {G.}~\bibnamefont {Ban}}, \bibinfo {author}
  {\bibfnamefont {J.}~\bibnamefont {Br\'egeault}}, \bibinfo {author}
  {\bibfnamefont {G.}~\bibnamefont {Darius}}, \bibinfo {author} {\bibfnamefont
  {D.}~\bibnamefont {Durand}}, \bibinfo {author} {\bibfnamefont
  {X.}~\bibnamefont {Fl\'echard}}, \bibinfo {author} {\bibfnamefont
  {M.}~\bibnamefont {Herbane}}, \bibinfo {author} {\bibfnamefont
  {M.}~\bibnamefont {Labalme}}, \bibinfo {author} {\bibfnamefont
  {E.}~\bibnamefont {Li\'enard}}, \bibinfo {author} {\bibfnamefont
  {F.}~\bibnamefont {Mauger}}, \bibinfo {author} {\bibfnamefont
  {Y.}~\bibnamefont {Merrer}}, \bibinfo {author} {\bibfnamefont
  {O.}~\bibnamefont {Naviliat-Cuncic}}, \bibinfo {author} {\bibfnamefont
  {J.~C.}\ \bibnamefont {Thomas}}, \ and\ \bibinfo {author} {\bibfnamefont
  {C.}~\bibnamefont {Vandamme}},\ }\href@noop {} {\bibfield  {journal}
  {\bibinfo  {journal} {Nucl.\ Instrum.\ Methods A}\ }\textbf {\bibinfo
  {volume} {565}},\ \bibinfo {pages} {876} (\bibinfo {year}
  {2006})}\BibitemShut {NoStop}%
\bibitem [{\citenamefont {Ban}\ \emph {et~al.}(2013)\citenamefont {Ban},
  \citenamefont {Durand}, \citenamefont {Fl\'echard}, \citenamefont
  {Li\'enard},\ and\ \citenamefont {Naviliat-Cuncic}}]{Ban2013}%
  \BibitemOpen
  \bibfield  {author} {\bibinfo {author} {\bibfnamefont {G.}~\bibnamefont
  {Ban}}, \bibinfo {author} {\bibfnamefont {D.}~\bibnamefont {Durand}},
  \bibinfo {author} {\bibfnamefont {X.}~\bibnamefont {Fl\'echard}}, \bibinfo
  {author} {\bibfnamefont {E.}~\bibnamefont {Li\'enard}}, \ and\ \bibinfo
  {author} {\bibfnamefont {O.}~\bibnamefont {Naviliat-Cuncic}},\ }\href@noop {}
  {\bibfield  {journal} {\bibinfo  {journal} {Ann. Phys. (Berlin)}\ }\textbf
  {\bibinfo {volume} {525 (8-9)}},\ \bibinfo {pages} {576} (\bibinfo {year}
  {2013})}\BibitemShut {NoStop}%
\bibitem [{\citenamefont {Delahaye}(2015)}]{Dela2015}%
  \BibitemOpen
  \bibfield  {author} {\bibinfo {author} {\bibfnamefont {P.}~\bibnamefont
  {Delahaye}},\ }\href@noop {} {\bibfield  {journal} {\bibinfo  {journal}
  {Nucl. Instrum. Methods A}\ } (\bibinfo {year} {2015})}\BibitemShut {NoStop}%
\bibitem [{\citenamefont {Fl\'echard}\ \emph
  {et~al.}(2011{\natexlab{a}})\citenamefont {Fl\'echard}, \citenamefont
  {Velten}, \citenamefont {Li\'enard}, \citenamefont {M\'ery}, \citenamefont
  {Rodr\'iguez}, \citenamefont {Ban}, \citenamefont {Durand}, \citenamefont
  {Mauger}, \citenamefont {Naviliat-Cuncic},\ and\ \citenamefont
  {Thomas}}]{Flec2011}%
  \BibitemOpen
  \bibfield  {author} {\bibinfo {author} {\bibfnamefont {X.}~\bibnamefont
  {Fl\'echard}}, \bibinfo {author} {\bibfnamefont {P.}~\bibnamefont {Velten}},
  \bibinfo {author} {\bibfnamefont {E.}~\bibnamefont {Li\'enard}}, \bibinfo
  {author} {\bibfnamefont {A.}~\bibnamefont {M\'ery}}, \bibinfo {author}
  {\bibfnamefont {D.}~\bibnamefont {Rodr\'iguez}}, \bibinfo {author}
  {\bibfnamefont {G.}~\bibnamefont {Ban}}, \bibinfo {author} {\bibfnamefont
  {D.}~\bibnamefont {Durand}}, \bibinfo {author} {\bibfnamefont
  {F.}~\bibnamefont {Mauger}}, \bibinfo {author} {\bibfnamefont
  {O.}~\bibnamefont {Naviliat-Cuncic}}, \ and\ \bibinfo {author} {\bibfnamefont
  {J.~C.}\ \bibnamefont {Thomas}},\ }\href@noop {} {\bibfield  {journal}
  {\bibinfo  {journal} {J. Phys. G: Nucl. Part. Phys.}\ }\textbf {\bibinfo
  {volume} {38}},\ \bibinfo {pages} {055101} (\bibinfo {year}
  {2011}{\natexlab{a}})}\BibitemShut {NoStop}%
\bibitem [{\citenamefont {Couratin}\ \emph {et~al.}(2012)\citenamefont
  {Couratin}, \citenamefont {Velten}, \citenamefont {Fl\'echard}, \citenamefont
  {Li\'enard}, \citenamefont {Ban}, \citenamefont {Cassimi}, \citenamefont
  {Delahaye}, \citenamefont {Durand}, \citenamefont {Hennecart}, \citenamefont
  {Mauger}, \citenamefont {M\'ery}, \citenamefont {Naviliat-Cuncic},
  \citenamefont {Patyk}, \citenamefont {Rodr\'iguez}, \citenamefont
  {Siegie\'n-Iwaniuk},\ and\ \citenamefont {Thomas}}]{Cour2012}%
  \BibitemOpen
  \bibfield  {author} {\bibinfo {author} {\bibfnamefont {C.}~\bibnamefont
  {Couratin}}, \bibinfo {author} {\bibfnamefont {P.}~\bibnamefont {Velten}},
  \bibinfo {author} {\bibfnamefont {X.}~\bibnamefont {Fl\'echard}}, \bibinfo
  {author} {\bibfnamefont {E.}~\bibnamefont {Li\'enard}}, \bibinfo {author}
  {\bibfnamefont {G.}~\bibnamefont {Ban}}, \bibinfo {author} {\bibfnamefont
  {A.}~\bibnamefont {Cassimi}}, \bibinfo {author} {\bibfnamefont
  {P.}~\bibnamefont {Delahaye}}, \bibinfo {author} {\bibfnamefont
  {D.}~\bibnamefont {Durand}}, \bibinfo {author} {\bibfnamefont
  {D.}~\bibnamefont {Hennecart}}, \bibinfo {author} {\bibfnamefont
  {F.}~\bibnamefont {Mauger}}, \bibinfo {author} {\bibfnamefont
  {A.}~\bibnamefont {M\'ery}}, \bibinfo {author} {\bibfnamefont
  {O.}~\bibnamefont {Naviliat-Cuncic}}, \bibinfo {author} {\bibfnamefont
  {Z.}~\bibnamefont {Patyk}}, \bibinfo {author} {\bibfnamefont
  {D.}~\bibnamefont {Rodr\'iguez}}, \bibinfo {author} {\bibfnamefont
  {K.}~\bibnamefont {Siegie\'n-Iwaniuk}}, \ and\ \bibinfo {author}
  {\bibfnamefont {J.~C.}\ \bibnamefont {Thomas}},\ }\href@noop {} {\bibfield
  {journal} {\bibinfo  {journal} {Phys.\ Rev.\ Lett.}\ }\textbf {\bibinfo
  {volume} {108}},\ \bibinfo {pages} {243201} (\bibinfo {year}
  {2012})}\BibitemShut {NoStop}%
\bibitem [{\citenamefont {Couratin}\ \emph {et~al.}(2013)\citenamefont
  {Couratin}, \citenamefont {Fabian}, \citenamefont {Fabre}, \citenamefont
  {Pons}, \citenamefont {Fl\'echard}, \citenamefont {Li\'enard}, \citenamefont
  {Ban}, \citenamefont {Breitenfeldt}, \citenamefont {Delahaye}, \citenamefont
  {Durand}, \citenamefont {M\'ery}, \citenamefont {Naviliat-Cuncic},
  \citenamefont {Porobic}, \citenamefont {Qu\'em\'ener}, \citenamefont
  {Rodr\'iguez}, \citenamefont {Severijns}, \citenamefont {Thomas},\ and\
  \citenamefont {Gorp}}]{Cour2013}%
  \BibitemOpen
  \bibfield  {author} {\bibinfo {author} {\bibfnamefont {C.}~\bibnamefont
  {Couratin}}, \bibinfo {author} {\bibfnamefont {X.}~\bibnamefont {Fabian}},
  \bibinfo {author} {\bibfnamefont {B.}~\bibnamefont {Fabre}}, \bibinfo
  {author} {\bibfnamefont {B.}~\bibnamefont {Pons}}, \bibinfo {author}
  {\bibfnamefont {X.}~\bibnamefont {Fl\'echard}}, \bibinfo {author}
  {\bibfnamefont {E.}~\bibnamefont {Li\'enard}}, \bibinfo {author}
  {\bibfnamefont {G.}~\bibnamefont {Ban}}, \bibinfo {author} {\bibfnamefont
  {M.}~\bibnamefont {Breitenfeldt}}, \bibinfo {author} {\bibfnamefont
  {P.}~\bibnamefont {Delahaye}}, \bibinfo {author} {\bibfnamefont
  {D.}~\bibnamefont {Durand}}, \bibinfo {author} {\bibfnamefont
  {A.}~\bibnamefont {M\'ery}}, \bibinfo {author} {\bibfnamefont
  {O.}~\bibnamefont {Naviliat-Cuncic}}, \bibinfo {author} {\bibfnamefont
  {T.}~\bibnamefont {Porobic}}, \bibinfo {author} {\bibfnamefont
  {G.}~\bibnamefont {Qu\'em\'ener}}, \bibinfo {author} {\bibfnamefont
  {D.}~\bibnamefont {Rodr\'iguez}}, \bibinfo {author} {\bibfnamefont
  {N.}~\bibnamefont {Severijns}}, \bibinfo {author} {\bibfnamefont {J.-C.}\
  \bibnamefont {Thomas}}, \ and\ \bibinfo {author} {\bibfnamefont {S.~V.}\
  \bibnamefont {Gorp}},\ }\href@noop {} {\bibfield  {journal} {\bibinfo
  {journal} {Phys.\ Rev.\ A}\ }\textbf {\bibinfo {volume} {88}},\ \bibinfo
  {pages} {041403(R)} (\bibinfo {year} {2013})}\BibitemShut {NoStop}%
\bibitem [{\citenamefont {Naviliat-Cuncic}\ and\ \citenamefont
  {severijns}(2009)}]{Navi2009}%
  \BibitemOpen
  \bibfield  {author} {\bibinfo {author} {\bibfnamefont {O.}~\bibnamefont
  {Naviliat-Cuncic}}\ and\ \bibinfo {author} {\bibfnamefont {N.}~\bibnamefont
  {severijns}},\ }\href@noop {} {\bibfield  {journal} {\bibinfo  {journal}
  {Phys. Rev. Lett.}\ }\textbf {\bibinfo {volume} {102}},\ \bibinfo {pages}
  {142302} (\bibinfo {year} {2009})}\BibitemShut {NoStop}%
\bibitem [{\citenamefont {Cornejo}, \citenamefont {Escobedo},\ and\
  \citenamefont {Rodr\'iguez}(2014)}]{Corn2014}%
  \BibitemOpen
  \bibfield  {author} {\bibinfo {author} {\bibfnamefont {J.~M.}\ \bibnamefont
  {Cornejo}}, \bibinfo {author} {\bibfnamefont {P.}~\bibnamefont {Escobedo}}, \
  and\ \bibinfo {author} {\bibfnamefont {D.}~\bibnamefont {Rodr\'iguez}},\
  }\href@noop {} {\bibfield  {journal} {\bibinfo  {journal} {Hyperfine
  Interact.}\ }\textbf {\bibinfo {volume} {227}},\ \bibinfo {pages} {223}
  (\bibinfo {year} {2014})}\BibitemShut {NoStop}%
\bibitem [{\citenamefont {Escobedo}(2014)}]{Esco2014}%
  \BibitemOpen
  \bibfield  {author} {\bibinfo {author} {\bibfnamefont {P.}~\bibnamefont
  {Escobedo}},\ }\emph {\bibinfo {title} {Desarrollo de un sistema de control
  para l\'aseres de diodo utilizando moduladores ac\'ustico-\'opticos}},\
  \href@noop {} {\bibinfo {type} {Master's thesis}},\ \bibinfo  {school}
  {Universidad de Granada} (\bibinfo {year} {2014})\BibitemShut {NoStop}%
\bibitem [{\citenamefont {Dawson}(1995)}]{Daw1995}%
  \BibitemOpen
  \bibfield  {author} {\bibinfo {author} {\bibfnamefont {P.~H.}\ \bibnamefont
  {Dawson}},\ }\href@noop {} {\emph {\bibinfo {title} {Quadrupole Mass
  Spectrometry and its Applications}}}\ (\bibinfo  {publisher} {American
  Institute for Physics, Elsevier Scientiﬁc Publishing Company, New York},\
  \bibinfo {year} {1995})\BibitemShut {NoStop}%
\bibitem [{\citenamefont {Olivero}\ and\ \citenamefont
  {Longbothum}(1977)}]{Oliv1977}%
  \BibitemOpen
  \bibfield  {author} {\bibinfo {author} {\bibfnamefont {J.}~\bibnamefont
  {Olivero}}\ and\ \bibinfo {author} {\bibfnamefont {R.}~\bibnamefont
  {Longbothum}},\ }\href@noop {} {\bibfield  {journal} {\bibinfo  {journal} {J.
  Quant. Spectrosc. Radiat. Transfer}\ }\textbf {\bibinfo {volume} {17}},\
  \bibinfo {pages} {233} (\bibinfo {year} {1977})}\BibitemShut {NoStop}%
\bibitem [{\citenamefont {Fl\'echard}\ \emph
  {et~al.}(2011{\natexlab{b}})\citenamefont {Fl\'echard}, \citenamefont {Ban},
  \citenamefont {Durand}, \citenamefont {Li\'enard}, \citenamefont {Mauger},
  \citenamefont {M\'ery}, \citenamefont {Naviliat-Cuncic}, \citenamefont
  {Rodr\'iguez}, \citenamefont {Thomas},\ and\ \citenamefont
  {Velten}}]{Flec2010}%
  \BibitemOpen
  \bibfield  {author} {\bibinfo {author} {\bibfnamefont {X.}~\bibnamefont
  {Fl\'echard}}, \bibinfo {author} {\bibfnamefont {G.}~\bibnamefont {Ban}},
  \bibinfo {author} {\bibfnamefont {D.}~\bibnamefont {Durand}}, \bibinfo
  {author} {\bibfnamefont {E.}~\bibnamefont {Li\'enard}}, \bibinfo {author}
  {\bibfnamefont {F.}~\bibnamefont {Mauger}}, \bibinfo {author} {\bibfnamefont
  {A.}~\bibnamefont {M\'ery}}, \bibinfo {author} {\bibfnamefont
  {O.}~\bibnamefont {Naviliat-Cuncic}}, \bibinfo {author} {\bibfnamefont
  {D.}~\bibnamefont {Rodr\'iguez}}, \bibinfo {author} {\bibfnamefont {J.~C.}\
  \bibnamefont {Thomas}}, \ and\ \bibinfo {author} {\bibfnamefont
  {P.}~\bibnamefont {Velten}},\ }\href@noop {} {\bibfield  {journal} {\bibinfo
  {journal} {Hyperfine Interact.}\ }\textbf {\bibinfo {volume} {199}},\
  \bibinfo {pages} {21} (\bibinfo {year} {2011}{\natexlab{b}})}\BibitemShut
  {NoStop}%
\bibitem [{\citenamefont {Ruiz}(2014)}]{Ruiz2014}%
  \BibitemOpen
  \bibfield  {author} {\bibinfo {author} {\bibfnamefont {E.}~\bibnamefont
  {Ruiz}},\ }\emph {\bibinfo {title} {Desarrollo de una microtrampa para
  experimentos de precisi\'on con iones de $^{40}$Ca}},\ \href@noop {}
  {\bibinfo {type} {Bachelor's thesis}},\ \bibinfo  {school} {Universidad de
  Granada} (\bibinfo {year} {2014})\BibitemShut {NoStop}%
\end{thebibliography}%

\end{document}